\documentclass[]{informs3}

\OneAndAHalfSpacedXI

\usepackage{natbib}
 \bibpunct[, ]{(}{)}{,}{a}{}{,}%
 \def\bibfont{\small}%
 \def\bibsep{\smallskipamount}%
 \def\bibhang{24pt}%
\TheoremsNumberedThrough     
\ECRepeatTheorems

\EquationsNumberedThrough    

\MANUSCRIPTNO{}

\usepackage{algorithm}
\usepackage{algpseudocode} 
\usepackage{float}
\usepackage{ifthen}

\def\masterCommentsSwitch{0}

\def\redTextAsRed{\masterCommentsSwitch}            	
\def\blueTextAsBlue{\masterCommentsSwitch}          	
\def\greenTextAsGreen{\masterCommentsSwitch}        	
\def\cyanTextAsCyan{\masterCommentsSwitch}		    	
\def\magentaTextAsMagenta{\masterCommentsSwitch}    	

\def\commentsOn{\masterCommentsSwitch}                 
\def\defaultCommentColour{olive}  

\newcommand{\RedText}[1]{\ifthenelse{\redTextAsRed = 1}{{\color{red} #1}}{#1}}
\newcommand{\BlueText}[1]{\ifthenelse{\blueTextAsBlue = 1}{{\color{blue} #1}}{#1}}
\newcommand{\GreenText}[1]{\ifthenelse{\greenTextAsGreen = 1}{{\color{green} #1}}{#1}}
\newcommand{\MagentaText}[1]{\ifthenelse{\magentaTextAsMagenta = 1}{{\color{magenta} #1}}{#1}}
\newcommand{\CyanText}[1]{\ifthenelse{\cyanTextAsCyan = 1}{{\color{cyan} #1}}{#1}}

\newcommand{\CommentText}[1]{
    \ifthenelse{\commentsOn = 1}{{\color{\defaultCommentColour}{\noindent [{\bf Comment:} #1]}}}{}
}

\newcommand{\CommentTextN}[2]{
    \ifthenelse{\commentsOn = 1}{{\color{\defaultCommentColour}{\noindent [{\bf Comment (#1):} #2]}}}{}
}

\newcommand{\CommentTextC}[2]{
    \ifthenelse{\commentsOn = 1}{{\color{#1}{\noindent [{\bf Comment:} #2]}}}{}
}

\newcommand{\CommentTextCN}[3]{
    \ifthenelse{\commentsOn = 1}{{\color{#1}{\noindent [{\bf Comment (#2):} #3]}}}{}
}

\newcommand{\CommentTextLong}[1]{
    \ifthenelse{\commentsOn = 1}{{\color{\defaultCommentColour}{\noindent [{\bf --- Comment BEGIN ---}\newline\noindent #1 \ \newline\hspace*{\fill}{\bf --- Comment END ---}]}}}{}
}

\newcommand{\CommentTextLongN}[2]{
    \ifthenelse{\commentsOn = 1}{{\color{\defaultCommentColour}{\noindent [{\bf --- Comment (#1) BEGIN ---}\newline\noindent #2 \ \newline\hspace*{\fill}{\bf --- Comment (#1) END ---}]}}}{}
}

\newcommand{\CommentTextLongC}[2]{
    \ifthenelse{\commentsOn = 1}{{\color{#1}{\noindent [{\bf --- Comment BEGIN ---}\newline\noindent #2 \ \newline\hspace*{\fill}{\bf --- Comment END ---}]}}}{}
}

\newcommand{\CommentTextLongCN}[3]{
    \ifthenelse{\commentsOn = 1}{{\color{#1}{\noindent [{\bf --- Comment (#2) BEGIN ---}\newline\noindent #3 \ \newline\hspace*{\fill}{\bf --- Comment (#2) END ---}]}}}{}
}

\newcommand{\Real}{\mathbb{R}}
\newcommand{\Rn}[1]{\mathbb{R}^{#1}}

\newcommand{\Prob}[1]{\mathbb{P}\left(#1\right)}

\usepackage{enumerate}
\usepackage{tikz}
\usetikzlibrary{shapes}
\usetikzlibrary{arrows}
\usetikzlibrary{plotmarks}

\usepackage{xr}
\begin{document}


\RUNAUTHOR{Chen}

\RUNTITLE{Social Networks and Choice}

\TITLE{Social Networks and the Choices People Make}

\ARTICLEAUTHORS{%
	\AUTHOR{Jeremy Chen}
		\AFF{Department of Decision Sciences, National University of Singapore Business School\\15 Kent Ridge Drive, Singapore 119245,
			\EMAIL{jeremy.chen@nus.edu.sg}
		}
} 


\ABSTRACT{%
Social marketing is becoming increasingly important in contemporary business. Central to social marketing is quantifying how consumers choose between alternatives and how they influence each other.
This work considers a new but simple multinomial choice model for multiple agents connected in a recommendation network based on the explicit modeling of choice adoption behavior.
Efficiently computable closed-form solutions, absent from analyses of threshold/cascade models, are obtained together with insights on how the network affects aggregate decision making.
In particular, a new measure of the overall decision making power of individual agents, ``decision share'', is proposed.
A stylized ``brand ambassador'' selection problem is posed to model targeting in social marketing. Therein, it is shown that a greedy selection strategy leads to solutions achieving at least $1-1/e$ of the optimal value.
In an extended example of imposing exogenous controls, a pricing problem is considered wherein it is shown that the single player profit optimization problem is concave, implying the existence of pure strategy equilibria for the associated pricing game. 
}%


\HISTORY{Working Paper. (Last Updated: \today)}

\maketitle

%



\section{Introduction}

People seldom make choices in isolation.
Not only are they influenced by the recommendations of prominent public figures \citep{Chung2011, Grover2011}, people are also influenced by family, friends and others in the communities or interest groups they belong to.

While businesses have long been aware of both recommendation effects, celebrity endorsements have historically been the most prominent mode of recommendation-based persuasion.
Over the past decade, however, improvements in communications technology and increased access to that technology have greatly facilitated the business use of peer-to-peer recommendation systems to drive sales.
Numerous web-based communities have emerged, allowing consumers to share their views and experiences on products/services, and engage in discussion on those views.

A growing body of evidence, much drawn from the aforementioned web-based communities, strongly suggests that consumers are highly responsive to recommendations from people within their social circles (see, for instance, 
\citealp{Keller2003, Smith2005, Bart2005, Bell2007, Iyengar2009, Iyengar2010, Racherla2012, eMarketerFeb2014}).
As such firms have been increasingly seeking to co-opt consumers as product evangelists.

Even in retail, social marketing is gaining prominence. Apple
and many other large firms use the ``Net Promoter System'' \citep{netpromotersystem2013} where responses to questions like ``{\em how likely would you be to recommend...}'' are aggregated to give an estimate of the difference between the percentage of ``promoters'' and the percentage of ``detractors''.

Internet social networks have stepped up to monetize the online social interaction that they mediate.
For instance, the advertising system of Facebook, the largest social networking site of the Western world, is premised on making (targeted) advertisements more compelling through endorsements (``likes'') by friends. As a testament to advertisers' increased focus on social marketing and their buy-in to the logic of Facebook's advertising scheme,
in 2011, Facebook was already serving up about \$1.58 billion worth of Internet display advertisements in the United States (\$3.15 billion worldwide). This figure rose to \$2.07 billon (\$4.28 billion worldwide) in 2012 and \$3.17 billion (\$6.99 billion worldwide) in 2013.

This work is premised on the proposition that choice is not only driven by personal preferences and expertise, but also by the recommendations and choices of others.
Here, how multiple agents, connected in a recommendation network, choose between alternatives is studied. Here, agents face the same finite set of alternative choices and possess idiosyncratic inclinations towards those choices.
However, as they simultaneously influence and are influenced by each other, the actual choices they make typically would differ from what would be suggested purely based on those inclinations.


The main contributions of this work are as follows:
\begin{enumerate}[(i)]
	\item
		This work presents a simple choice model that explicitly models the key feature of recommendation networks: the agents' option to adopt other agents' choices.
	\item
		Analytical solutions are readily obtained from the model, providing insights on how a recommendation network affects choice. This characteristic has not been demonstrated in existing threshold/cascade models (wherein simulation is generally required to compute solutions).
	\item
		Complementing the usual ``market share'' (``choice share'' here) quantifications, a natural measure of the influence of each agent (``decision share'') is developed. This measure also turns out to be the key to the natural generalization beyond finite choice sets.
	\item
		The ``brand ambassador'' selection problem is presented to model targeting in social marketing. Though it is NP-hard, $(1-1/e)$-optimal solutions can be efficiently computed.
\end{enumerate}


\subsection{Relationship With Existing Work}

	A number of authors have studied choice models where membership in social groups affects outcomes.
Notable examples are the work of \cite{Bramoulle2009} and \cite{DeGiorgi2010}, who studied model identification in a setting where decisions of agents (activity levels and binary choice respectively) are affine in the ``average behavior'' of the agents' social groups (``linear in means''); \cite{Brock2001} who considered (binary) random utility maximization models where agents' utilities are affine in the ``average group behavior''; and \cite{Brock2002} who extended the aforementioned binary model to a multinomial choice model wherein the utility accruing to an agent for selecting a given choice is affine in the average choice probability for that choice.
Unfortunately, as \cite{Blume2010} note in their survey article, there exists little work on multinomial choice models with social interactions.
Furthermore, closed form solutions are generally not available even for binomial choice models.
In this work, a new multinomial choice model that models the effects of a recommendation network is introduced. The model is a variant of the ``linear in means'' model but generates (closed form) choice probabilities rather than ``activity levels''. As a bonus, the model is readily generalizable to infinite choice sets.

	Also relevant to this work is the research program on learning in networks.
The primary project within this domain of inquiry is the characterization of when beliefs do or do not converge to (the relevant representation of) true state of the world.
The literature on learning in networks may be loosely classified into two sub-schools: Bayesian learning (e.g.: \citealp{Acemoglu2011}), and ``linear'' learning (e.g.: \citealp{DeGroot1974,Acemoglu2010,Acemoglu2013}).
In this paper, a close connection will be drawn between this work and a general model of ``linear'' learning.


	Viewed from the angle of social media marketing, this work has links to the literature on the propagation of information/influence on a network.
The influence maximization problem was first proposed and studied by \cite{Domingos2001}. In that work, they proposed a general descriptive model of influence propagation based on a Markov random field, but proposed only simple local search heuristics for solving the influence maximization problem.
Shortly after, \cite{Kempe2003,Kempe2005} considered two families of models of influence propagation, ``cascade'' models and ``threshold'' models, and applied them to influence maximization.
By demonstrating that the expected number of ``active nodes'' on termination of a given influence propagation process is monotone and submodular in the initial subset of ``active nodes'', they obtained $(1-1/e-\epsilon)$-optimal ($\epsilon$ arbitrarily small, according to the fidelity of the simulation used) performance guarantees for greedy selection strategies in the influence maximization problem under the given influence propagation process. Notably, the natural ``cascade'' process generating the same ``choice shares'' as the model to be presented does not satisfy the assumptions of the ``decreasing cascade model'', the most general cascade model analyzed in \cite{Kempe2005} (see \textsection\ref{ss_comparison_KKT}).

	Distinct from previous work, underpinning the model is an implicit extension of the choice set from the base set of alternatives to include the adoption of other agents' choices, underscoring the fact that adopting the choice of a particular agent is itself a distinct choice and reflecting a key aspect of recommendation networks.
Closed form choice probabilities may be obtained in a computational tractable manner, making the model practical for large-scale social marketing applications.
Like \cite{Kempe2003,Kempe2005},
$(1-1/e)$-optimal solutions to the ``brand ambassador'' selection problem may be obtained via the route of monotonicity and submodularity, arguably fundamental characteristics of the diffusion of information on a network.


\section{Networks and Choice Modeling: Fundamentals}


A discrete choice model is a (stochastic) decision model wherein agents are modeled as using some given decision processes, such as solving optimization problems, to arrive at their choices.
However, due to imperfect knowledge on the part of the modeler, there is uncertainty (e.g.: parametric) about the agents' decision processes, and thus the choices that will be made. As such, choice models return not a single choice for each agent, but a probability distribution over the set of alternatives.
The most commonly used framework, random utility models, has agents maximizing over a (finite) set of random utility functions, encompassing models like the logit, probit and mixed logit.
(See \cite{Train2009} for an introduction, and see \cite{Natarajan2009} for an interesting connection to discrete optimization under uncertainty.)
\BlueText{More broadly, discrete choice models feature stochastic decision dynamics being modeled at varying levels of detail. For example, \cite{Blanchet2013} introduce structure by modeling substitution between choices as state transitions of a Markov chain; and at a distant extreme, \cite{Farias2013} completely eschew modeling decision dynamics in favor of estimating distributions over all possible preference orders.}


In this paper, a choice model is presented based on a simple probabilistic decision process: agents probabilistically make choices based on their own preferences or adopt the choices of another agent.

Prior to describing the model, it would be useful to first define the nomenclature:
\begin{itemize}
	\item
		$A$: The (finite) set of agents.
	\item
		$C$: The (finite) set of choices.
	\item
		$p_{ik} \in [0,1]$: Probability that agent $i\in A$ will adopt the choice of agent $k\in A$.
	\item
		$P$: A matrix of dimension $|A|\times|A|$ with $(i,k)$ entry $p_{ik}$.
	\item
		$q_{ij} \in [0,1]$: Probability that agent $i\in A$ will select choice $j\in C$ without consulting the network.
	\item
		$q^{(j)}$: A vector of dimension $|A|$ with $i$-th entry $q_{ij}$.
	\item
		$\pi_{ij}$: Probability that agent $i\in A$ will pick choice $j\in C$.
\end{itemize}
Presently, the $p_{ik}$'s and the $q_{ij}$'s are defined as constant parameters. However, $P$ and $\{q^{(j)}\}_{j\in C}$ will be allowed to vary in the ``brand ambassador'' selection problem of \textsection\ref{ss_influence_max} and in the pricing problem of \textsection\ref{ss_optimization}. Subsequently, let $e_i$ be the $i$-th unit vector and let $e$ be the vector of all ones.

Though the estimation of model parameters is important, the focus of this work is modeling rather than inference. Furthermore, the right estimation process depends on the available data. Nevertheless, a simple approach will be presented in
Appendix \ref{appendix_estimation} of the online supplement.


\subsection{The Basic Model}
\label{ss_model}

	The model considers the $\pi_{ij}$'s to be ``steady state'' choice probabilities which allows one to dispense with the (likely contentious) stipulation of decision making dynamics. At the core of this choice model is the explicit modeling of a key aspect of recommendation networks, agents adopting the choices of other agents. The choice probabilities may then be described
as follows for $i\in A$, $j\in C$:
	\begin{eqnarray}
		\pi_{ij} & =	& \Prob{i{\ \rm chooses\ }j,\ i{\ \rm does\ not\ consult\ network}} \nonumber\\
				&    & +\ \Prob{i{\ \rm adopts\ some\ other\ agent's\ choice\ of\ }j} \label{choice_model_0} \\
				& {\rm modeled}\atop{=}
					& \Prob{i{\ \rm chooses\ }j,\ i{\ \rm does\ not\ consult\ network}} \nonumber\\
				& & +\ \sum_{k\in A}\Prob{i{\ \rm adopts\ }k{\rm 's\ choice}}\Prob{k{\ \rm chooses\ }j} \label{choice_model_1} \\
				& = & q_{ij} + \sum_{k \in A}{p_{ik} \pi_{kj}} \label{choice_model_2}.
	\end{eqnarray}

	In this model, agents' actions are driven by other agents' actions, not their expectations of other agents' actions.
(One might make a consistency argument that both should coincide in ``steady state''.)
This may arise through a mechanism like observational learning, but that detail is not addressed here.
	Naturally, it is required that the probabilities for the various actions that $i$ may take sum to one. Therefore, for all $i\in A$,
	\begin{equation}
		\label{partition_of_activity}
		\sum_{k \in A}{p_{ik}} + \sum_{j \in C}{q_{ij}}  = 1,
	\end{equation}
	neatly segregating ``adoption'' and ``direct selection'' behavior.
	In addition, it is assumed that
	\begin{equation}
		p_{ii} = 0
	\end{equation}
	for all $i\in A$. Thus, each agent $i$ is allowed $|A|+|C|-1$ distinct actions: selecting some element of $C$ without consulting his/her network and choosing to adopt the choice of some other agent in $|A|$.

	\begin{example}[The Classical (Disconnected) Setting]
	When $P=0$, the choices agents make are not coupled through a recommendation network. So $\pi_{ij}=q_{ij}$ for all $i\in A$, $j\in C$.\hfill$\square$
	\end{example}

	The first term on the right hand side of equation (\ref{choice_model_0}) is the contribution from agent $i$ acting like an isolated individual, the second term accounts for agent $i$'s actions due to ``influence'' from other agents.
	Equation (\ref{choice_model_1}) encodes the modeling assumption that an agents adopts others' choices independently of the choices the latter make.
As such, the rejection of choices due to differing preferences is not modeled. Also, the phenomena whereby the knowledge of one's influence changes the choices one makes (image shaping) cannot be captured.
These modeling gaps are acknowledged.

	On a similar note, one implication of the model is that an agent may select, with non-zero probability, something that he/she would never choose if the agent were isolated. This is not an unrealistic outcome. An agent may accept the recommendations of other agents on the basis that they have information he/she does not have. In day to day life, people regularly accept recommendations from friends to ``try something new''. As such, this would just be an instance of an agent extending his/her choice consideration set due to peer influence.



	From a broader perspective, equation (\ref{partition_of_activity}) points to the implicit extension of the choice set from $C$ to include adopting advice from other agents, hinting at some underlying choice model over agents' actions.
In this light, this model might be thought of as a ``meta-model'' of choice into which single-agent choice models may be ``plugged''.


\subsection{Quantifying Choice Probabilities}\label{ss_choice_prob}

The following regularity condition is assumed to ensure the $\pi_{ij}$'s are well-defined:

	\begin{assumption}[Collective Decisiveness]
	\label{assumption}
	The set $Q:=\{i\in A: \sum_{j\in C}{q_{ij}} > 0 \} \not= \emptyset$, and for each $i\in A\backslash Q$, there exists a sequence of agents $(a_1^i, a_2^i, \ldots, a_{n_i+1}^i)$ such that $a_1^i = i$, $a_{n_i+1}^{i} \in Q$, and for $k=1,2,\ldots, n_i$, $p_{a_k^i a_{k+1}^i}>0$.
	\end{assumption}

	Assumption \ref{assumption} may be read as the requirement that, for each agent, there exists a ``probabilistic path'' leading from that agent to some choice.
	It is necessary and sufficient for choice probabilities to be well-defined, in a sense.
	The technical implications of Assumption \ref{assumption} are listed below:
	\begin{lemma}\label{M_well_defined_etc}
	The following hold if and only if Assumption \ref{assumption} holds:
	\begin{enumerate}[\ \ (a)]
		\item \label{spectral_radius_property}
			The spectral radius\footnote{The spectrum of a matrix $P$ is the set of its eigenvalues, denoted as $\Lambda(P)$. The spectral radius of a matrix $P$ is the maximum of the absolute value of the elements of $\Lambda(P)$. It is also equal to $\max\limits_{\|x\|_2=1}{\|Px\|_2}$ where $\|\cdot\|_2$ is the $2$-norm.} of $P$ (largest eigenvalue of $P$ by magnitude) is strictly less than unity.
		\item \label{M_well_defined_property}
			$(I-P)^{-1}$ is well-defined.
		\item \label{subM_well_defined_property}
			For any principal sub-matrix\footnote{Up to  permutations, all principle sub-matrices map bijectively to and from non-empty subsets of $A$. For any non-empty $B\subseteq A$, the principle sub-matrix corresponding to $B$ is the matrix where the rows and columns corresponding to elements in $A\backslash B$ are removed.}, $V$, of $P$, the matrix $(I-V)^{-1}$ is well-defined.
	\end{enumerate}
	\end{lemma}

	With that, based on the definition of the model, the choice probabilities may be computed through an elementary exercise in linear algebra. 

	\begin{proposition}[Individual Choice Probabilities]
	\label{choice_share}
	The probability of agent $i$ selecting choice $j$ is given by:
	\begin{equation}
		\pi_{ij} = e_i^{T}(I - P)^{-1}q^{(j)}. \label{choice_share_eq}
	\end{equation}
	\end{proposition}
	\proof{Proof.} The result follows from equations (\ref{choice_model_0})-(\ref{choice_model_2}) and part (\ref{M_well_defined_property}) of Lemma \ref{M_well_defined_etc}. \hfill$\blacksquare$
	\endproof

	To verify that, for agent $i\in A$, $\{\pi_{ij}\}_{j\in C}$ is indeed a probability distribution, observe that the non-negativity of $e_i$, $(I-P)^{-1}$ ($=I+P+P^2+\ldots$) and $q^{(j)}$ implies that $\pi_{ij}\ge 0$ ($j\in C$), and by recognizing that $\sum_{j \in C}{q^{(j)}}=e-Pe=(I-P)e$, it is easy to verify that $\sum_{j \in C}{\pi_{ij}} = 1$.

	\begin{example}[The Impact of Accepting Recommendations]\label{ex_influential_agent}
		Consider the situation where there is an ``influential'' agent (1) and two other agents (2 and 3).
		Suppose $C=\{A,B\}$, and
		$$
			P = \left[\begin{array}{ccc}
					0 & 1/8 & 1/8 \\
					1/2 & 0 & 1/4 \\
					1/2 & 1/4 & 0
				\end{array}\right],\ \ 
			q^{(A)} =
			\left[\begin{array}{c}
					1/2 \\
					0 \\
					0 
				\end{array}\right],\ \ 
			q^{(B)} =
			\left[\begin{array}{c}
					1/4 \\
					1/4 \\
					1/4
				\end{array}\right].
		$$		
		This is illustrated in Figure \ref{fig_example} wherein agents are represented by circles, choices are represented by squares, and decisions by agents are represented by outgoing arcs.
%
%
		\begin{figure}[!h]	
			\begin{center}
			\caption{A visual representation of Examples \ref{ex_influential_agent},  \ref{ex_influential_agent_choice_share} and \ref{ex_influential_agent_DS}.}
			\label{fig_example}
			\begin{tikzpicture}[->,>=stealth',shorten >=1pt,auto,node distance=3cm,thick,
				agent node/.style={circle,draw,font=\sffamily\bfseries\footnotesize},
				choice node/.style={rectangle,draw,font=\sffamily\footnotesize},
				every node/.style={font=\sffamily\scriptsize},
				x=0.70cm,y=0.70cm
				]
			
			  \node[agent node] (1) at (0,0) {1};
			  \node[choice node] (1A) at (-1,1.5) {A};
			  \node[choice node] (1B) at (1,1.5) {B};
			  \node[agent node] (2) at (-2,-2.5) {2};
			  \node[choice node] (2B) at (-4,-2.5) {B};
			  \node[agent node] (3) at (2,-2.5) {3};
			  \node[choice node] (3B) at (4,-2.5) {B};
			
			  \path[every node]
			    (1) edge [bend left] node[right] {1/8} (3)
			        edge [bend right] node[left] {1/8} (2)		        
			        edge node[below left] {1/2} (1A)		        
			        edge node[below right] {1/4} (1B)		        
			    (2) edge node [below right] {1/2} (1)
			        edge node {1/4} (3)
			        edge node [below] {1/4} (2B)		        
			    (3) edge node [below left] {1/2} (1)
			        edge [bend left] node {1/4} (2)
			        edge node [below] {1/4} (3B);		        
			\end{tikzpicture}
			\end{center}
		\end{figure}
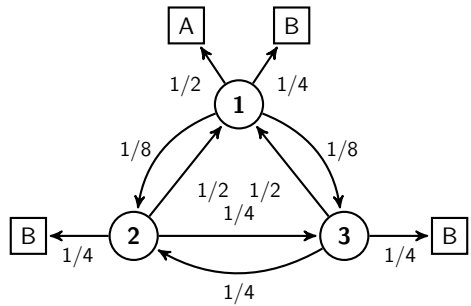
%
%
		In the absence of network effects (thus normalizing the $q^{(j)}$'s), agents 2 and 3 would each have chosen $B$ with probability $1$. Network effects, however, result in them choosing $A$ with probability $0.4$.\hfill$\square$
	\end{example}

	\RedText{It is useful to point out (again) that the outputs of the model are expectations. Unless decision dynamics are stipulated, the distribution of joint outcomes (vectors in $C^{|A|}$)
cannot be studied.
}


\subsection{Insights through Analogies: Decision Dynamics and Learning Dynamics}\label{ss_analogies}

	\BlueText{As with most network models, insights may be obtained from analogies. Subsequently, equation (\ref{choice_share_eq}) will be discussed from the perspectives of an appropriate discrete time Markov chain and also the ``linear learning'' of choice probabilities that both lead to equivalent probabilistic outcomes.
	}

	\subsubsection{The Markov Chain Analogy.} \label{sss_markovian_connection}

	\RedText{One may understand equation (\ref{choice_share_eq}) to be evaluating the limiting distribution (over absorbing states) of a random walk on a graph, describing the agents' adoption propensities and preferences over alternatives, given some initial distribution over its nodes (``states''). In this analogy, there are two types of states, ``agent $i$ to make the choice'' for each $i\in A$ and ``choice $j$ selected'' for each $j\in C$. The latter are, of course, absorbing states.
}

	In the ``first step'', agent $i$ chooses $j$ with probability $q_{ij}$ (``Choice $j$ selected'') and the probability that he adopts the choice of agent $k$ (``Agent $k$ to make the choice'') is $[e_i^{T} P]_k$. Reasoning along these lines, in the ``second step'', the probability of $j$ being picked increases by $e_i^{T} P q^{(j)}$; in the ``third step'', it increases by $e_i^{T} P^2 q^{(j)}$. This leads to the equivalent expression
$\pi_{ij} = e_i^{T}\left(I + P + P^2 + \ldots \right)q^{(j)}$.
Furthermore, under Assumption \ref{assumption}, with each step, the probability that a choice has not yet been made (the probability of the states $\{{\rm Agent\ }i{\rm\ to\ make\ the\ choice}\}_{i\in A}$) decreases towards zero.

	Note that one should not interpret sample paths in this Markov chain as instances of decision making dynamics. Rather, this analogy is a (standard) tool to explain how a certain measure of ``centrality'' arises in network models (see \textsection\ref{ss_decision_share}).


	\subsubsection{The Linear Learning Connection.} \label{sss_linear_learning_connection}

	There is a compelling connection of this model to models of ``linear learning''. Consider a discrete time system where, in each time step, each agent updates his/her beliefs by combining some private signal with the current beliefs of other agents (including himself/herself). Denoting the belief of agent $i$ at time $t$ as $x_i^{(t)}$, and assuming that the beliefs may be represented as real values, this may be written as:
$$
	x_i^{(t+1)}=\alpha_i\sum_{k}{v_{ik} x_k^{(t)}}+(1-\alpha_i)x_i^{(0)}
$$
where it is assumed that $\sum_{k}{v_{ik}}=1$ for all $i$, the $v_{ik}$'s are all non-negative and $\alpha_i\in[0,1]$ for all $i$.

	This is a generalization of the well-known DeGroot Model of linear learning \citep{DeGroot1974} (wherein $\alpha_i=1$ for all $i$) and is similar to what was presented by \cite{Acemoglu2010} (who studied a particular model of interaction and learning in a community). In contrast to the DeGroot Model, the above generalization enables agents to balance ``local'' information, represented by the agents' initial beliefs, with a ``global overview'' gained from aggregating information from sources across the network, and is thus a useful model of interaction and learning in communities.

	Under mild conditions analogous to Assumption \ref{assumption}, the vector of beliefs converges:
$$
	\lim\limits_{t\rightarrow\infty}\ x^{(t)}=(I-DV)^{-1}(I-D)x^{(0)}
$$
where $x^{(t)}$ is a vector of the $x_i^{(t)}$'s, $V$ is a matrix of the $v_{ik}$'s and $D$ is a diagonal matrix of the $\alpha_i$'s.

	Comparing this with equation (\ref{choice_share_eq}), $DV$ plays a similar role as $P$ and $(I-D)x^{(0)}$ plays a similar role as $q^{(j)}$ (for some $j\in C$). ($V$ is a scaling of $P$ so the rows sum to $1$, and $D$ reverses that scaling.) Given $P$ and $\{q^{(j)}\}_{j\in C}$, one is able to find $D$, $V$ and a $x^{(0)}$ for each $j$ such that the outcome of the linear learning process generates the choice probabilities.
Specifically, $V$ is the matrix obtained from $P$ by normalizing its rows to sum to $1$, and $\alpha_i$ is the sum of entries in row $i$ of $P$. To obtain $x^{(0)}$ from $q^{(j)}$, once simply divides the $i$-th entry by $1-\alpha_i$.

	One may observe that, for a given $i\in A$, $\{q_{ij} / (1 - \sum_{k\in A}{p_{ik}})\}_{j\in C}$ ($=\{q_{ij} / (1 - \alpha_i)\}_{j\in C}$) is a discrete probability distribution over $C$. (This follows from equation (\ref{partition_of_activity}).)
These discrete distributions may be thought of as representing each agent's idiosyncratic preferences over the set of possible choices (denote these ``personal choice probabilities'').
\RedText{The choice probabilities of the model may be viewed as the limit of a set of linear learning processes, one for each choice $j\in C$. In particular, for choice $j\in C$, the ``private signals'' used are vectors of the ``personal choice probabilities'' of the various agents for choice $j$, obtained from normalizing $\{q_{ij}\}_{j\in C}$ for each agent $i\in A$ and constructing a vector of the entries corresponding to $j$.}
	
	\RedText{One may argue that such processes of iterated averaging are suspect because it is unclear what manner of object emerges from each iteration of averaging. (E.g.: when averaging subjective survey responses across participants.)}
	However, ``personal choice probabilities'' have objective economic meaning. Performing a weighted average is, precisely, mixing. If one were to synchronously run $|C|$ such processes, one for each $j\in C$, one would find that in each time step, the property of being a set of probability distributions (one for each agent) is preserved, in particular in the limit.

	This discussion hints once more, albeit more directly, that the model is a kind of ``meta-model'' of choice wherein existing choice models may be augmented with information about agents' propensities to adopt the choices of others to study how social networks affect the choices people make.


\subsection{Computational Issues.}\label{ss_computational_issues}

	Choice probabilities (and later, choice shares) can be computed in $O(|A|^3)$ floating point operations with the coefficient of the leading term being $2/3$ (see, for instance, \citealp{Trefethen1997}). Alternatively, because the matrix $I-P$ is diagonally dominant and typically sparse, efficient iterative methods may be applied (see, for instance, \citealp{Saad2003}).

	However, it is important to note that data generally contains noise, leading to considerations of the sensitivity of choice share to perturbations of $P$ and $\{q^{(j)}\}_{j\in C}$.
Notably, whichever sensitivity analysis approach is taken, qualitatively, the less ``decisive'' the agents are (the smaller the numbers $\{\sum_{j\in C}{q_{ij}}\}_{i\in A}$ are), the closer the spectral radius of $P$ is to $1$, the more sensitive the choice shares are to the input data, and the greater the care needed when interpreting computational results.


\section{Choice in Communities}


Having touched on the basics of the model from the elementary perspective of a single agent making a choice, these ideas will now be extended to consider the extension to communities. ``Choice share'', an analogue to ``market share'', will be presented along with a new measure of decision making power of agents which will be termed ''decision share''. Decision share has a clear economic interpretation and turns out to present a direct route for generalization to infinite choice sets.

\subsection{Choice Share}

	While equation (\ref{choice_share_eq}) considers the behavior of a single agent, applications typically require consideration of the aggregate choices of a community. Therefore, a natural generalization would be to consider each agent having a non-negative endowment to allocate to the choices, and also to denote the expected amount (of the total endowment) allocated to a given choice its ``choice share''. \RedText{(This may be thought of as ``market share''.)} For each agent $i\in A$, let $i$'s endowment be $w_i\ge 0$ and let $w$ be a vector of the $w_i$'s. Denote the choice share of $j\in C$ with respect to endowment $w$, $\pi_j^w$:
	\begin{definition}[Choice Share]
		\label{choice_share_w}
		The choice share of choice $j$ with respect to endowment $w$ ($w \ge 0$) is given by 
		$\pi_{j}^{w} := \sum_{i\in A}w_i \pi_{ij}$.
	\end{definition}

	Clearly,
	\begin{equation}
		\pi_{j}^{w} = w^{T}(I - P)^{-1}q^{(j)}. \label{choice_model_w_eq}
	\end{equation}
	Furthermore, $\pi_j^w \ge 0$ for all $j\in C$ and $\sum_{j\in C}{\pi_j^w}=\sum_{i\in A}{w_i}$ (the entire endowment is allocated).

	\begin{example}[Choice Share: Revisiting Example \ref{ex_influential_agent}]\label{ex_influential_agent_choice_share}
		Consider Example \ref{ex_influential_agent} with $w = \frac{1}{3}e$. The choice shares of $A$ and $B$ turn out to be $7/15$ and $8/15$ respectively. \hfill$\square$
	\end{example}


\subsection{Decision Share}\label{ss_decision_share}

	Choice share measures allocation of a community's total endowment to choices in $C$. A related question would be how much of the allocation of the total endowment to choices in $C$ is, in expectation, (ultimately) determined by a given agent $i$.
Given an endowment vector $w$, let $\delta_i^{w}$ denote the ``decision share'' of agent $i\in A$, and define it as the expected amount allocated to choices in $C$ due to agent $i$ selecting a choice without consulting his/her network.
Decision share may be quantified as follows:

	\begin{proposition}[Decision Share]
		\label{decision_share}
		The decision share of agent $i\in A$ is given by:
		\begin{equation}\label{eqn_decision_share}
			\delta_i^{w} = c_i^{w} \bar q_i
		\end{equation}
		where $w$ ($w\ge 0$) is an endowment, $c^{w} = w^{T}(I-P)^{-1}$ and $\bar q = \sum_{j\in C}{q^{(j)}}$.
	\end{proposition}
	\proof{Proof.} To derive the $\delta_i^{w}$'s, a instance of the model will be considered where the choice shares measure how much of the total endowment each agent directly allocates to the choices in $C$.
Consider a problem with a new set of choices $\bar C = A$. Replace $\{q_{ij}\}_{i\in A,j\in C}$ with $\{\tilde q\}_{i\in A, j\in \bar C}$ such that for $i\in A$, $\tilde q_{ii} = \bar q_i = 1 - \sum_{k\in A}{p_{ik}}$ and $\tilde q_{ij} = 0$ for $j\not= i$.
The interpretation is that when a choice in $C$ is selected, what is tracked is not how much of the endowment is allocated to that choice, but rather which agent made that decision.
Noting that $\tilde q^{(i)} = (1 - e_i^{T}Pe)e_i = (e_i^{T}(e - Pe))e_i = (e_i^{T} \bar q) e_i$, and $\delta_i^{w} = w^{T}(I-P)^{-1} e_i e_i^{T} \bar q = c_i^{w} \bar q_i$ the proof is complete. \hfill$\blacksquare$
	\endproof

	\begin{example}[Decision Share: Revisiting Example \ref{ex_influential_agent}]\label{ex_influential_agent_DS}
		 Consider Example \ref{ex_influential_agent} with $w = \frac{1}{3}e$. The decision shares for agents $1$, $2$ and $3$ are $7/10$, $3/20$ and $3/20$ respectively.
\hfill$\square$
	\end{example}

	\begin{example}[Decision Share: Isotropic Fully Connected Network]\label{ex_fully_connected_network}
		Suppose for $\rho\in [0,1)$, $p_{ik} = \rho/(|A|-1)$ for all $i,k \in A$, $i\not= k$, then
		$$
			\delta_i^w = \frac{|A|-1}{|A|-1+\rho} \left( (1-\rho) w_i + \frac{\rho}{|A|-1} w^{T}e \right)
		$$
		gives the resulting decision shares for endowment $w$. \hfill$\square$
	\end{example}

	\begin{example}[Decision Share: ``Hub and Spoke'' Network]\label{ex_fully_connected_network}
		Let agent $h \in A$ be the ``hub'' in a ``hub and spoke'' network.
		Suppose $\rho \in [0,1]$, $p_{ih} = \rho$ for all $i,k \in A$, $i\not= h$, and suppose the other $p_{ik}'s$ are $0$. Then
		$$
			\delta_i^w = \left\{
				\begin{array}{ll}
					(1-\rho)w_i + \rho w^{T}e & (i = h)\\
					(1-\rho)w_i & (i \not= h)\\
				\end{array}
			\right.
		$$
		gives the resulting decision shares for endowment $w$. \hfill$\square$
	\end{example}

	\begin{example}[Decision Share: Fully Connected Network with a Hub]\label{ex_hub_agent}
		Let agent $h \in A$ be a ``hub'' in a fully connected network.
		Let $\rho_F \in [0,1), \rho_H \in (0,1]$ with $\rho_F + \rho_H \le 1$.
		Consider an increasing sequence of agents and their corresponding endowments $\{(A_n, w_{(n)})\}_{n\ge 1}$ with $|A_n|\rightarrow\infty$. For each $n\ge 1$, let $p_{ik} = \rho_F/(|A_n|-1)$ for all $i,k \in A_n$, $i\not= k$, $k\not= h$ and $p_{ih} = \rho_F/(|A_n|-1) + \rho_H$ for all $i \in A_n$, $i\not= h$.
		Suppose also that $|w_{(n),i} / w_{(n),k}| \le \Delta < \infty$ for all $i,k \in A_n$.
		Then,
		$$
			\lim_{n\rightarrow\infty}\frac{\delta_h^{w_{(n)}}}{w_{(n)}^{T}e} = \left(\frac{1}{\rho_H} - \frac{\rho_F}{1-\rho_F}\right)^{-1}.
		$$
		gives the asymptotic ratio of the decision share for the ``hub'' agent $h$ to the total endowment. \hfill$\square$
	\end{example}


	\RedText{The decision share of $i$ depends on the extent to which others in the network adopt $i$'s decisions, including choice adoption, ($c_i^{w}$) and how ``decisive'' $i$ is ($\bar q_i$). $c_i^{w}$ is a network characteristic best understood via the Markov chain analogy, and $\bar q_i$ is an individual characteristic measuring how often $i$ makes decisions without consulting the wider network. These bear further explanation.}

	\subsubsection{Centrality.}

	\RedText{In the discussion of \textsection\ref{sss_markovian_connection}, $[(I-P)^{-1}]_{ki}$ gave the expected frequency that state $i$ (``agent $i$ to make the choice'') would occur if the initial state were $k$ (``agent $k$ to make the choice''). With that analogy in mind, note that when $c_i^{w}$ ($=w^{T}(I-P)^{-1}e_i$) is large, agent $i$ would determine the (ultimate) choice more often. Thus, $c_i^{w}$ is a measure of the reliance of the network on agent $i$ for decision making. In particular, it is a weighted centrality measure for agent $i$, a generalization of Katz centrality \citep{Katz1953} or of Bonacich centrality \citep{Bonacich1987}.
}



	The ``centrality-connection'', pioneered by authors like \cite{Katz1953} and \cite{Bonacich1987}, has been made in many economic settings where agents interactions may be described using a network.
For example, in network games with local payoff complementarities, ``central'' agents benefit by free-riding on the efforts of other agents \citep{Ballester2006,Bramoulle2007}, or by being compensated for the positive externalities they exert \citep{Candogan2012}; In financial networks, shocks can be amplified through a network effect modulated by centrality and a market effect due to ``fire-sales'' of illiquid assets that further depress asset values \citep{Chen2013}.

	\subsubsection{Decisiveness.}
	\RedText{
	Yet, ``central'' as it has been in measuring the ``importance'' of nodes in networks, centrality is not the sole determinant of decision share. Decision share is the product of $c_i^{w}$ ($=w^{T}(I-P)^{-1}e_i$) and $\bar q_i$ ($=e_i^{T}(I-P)e$). The latter term is the $i$-th component of $\sum_{j\in C}{q^{(j)}}$, and the larger it is, the more likely agent $i$ makes a choice ``on his own'' rather than by adopting another agent's choice.
Thus, it is not unreasonable to use $\bar q_i$ as a measure of agent $i$'s ``decisiveness''.

	One might say that decision share is centrality --- the extent to which the community leans on an agent for decision making --- modulated by ``decisiveness'', which gives the likelihood that that agent ``decides'' when ``called upon'' to do so. This is reasonable and intuitive.

	Additionally, the more ``decisive'' agents in $A$ are, the lower the impact of network effects; and with ``indecisive agents'', one might reason that, given reasonable decision dynamics, ``herding'' would be observable in the joint distribution of choices. A brief discussion on this is provided in Appendix \ref{appendix_ex_decisiveness}, including the quantification of the expected size of the largest ``herd'' in a specialized setting which turns out to be a new result for the classical Polya urn model.
}

	\subsubsection{Putting the Two Together.}
\RedText{
	To connect the two, consider a mechanical analogy describing the ultimate selection of choices. Let there be a network of pipes with leaky joints, one corresponding to each $i\in A$. The fraction of fluid entering joint $i$ that flows out to joint $k\in A$ is $p_{ik}$ and the fraction that leaks is $\bar q_i$.
Flow from joint to joint corresponds to choice adoption, and leakage corresponds to ultimate choice selection.
Thus, giving joint $i$ an initial infusion of $w_i$ for each $i\in A$, the total flow into joint $i$ is $c_i^{w}=w^{T}(I-P)^{-1}e_i$. With a fraction $\bar q_i$ leaking out, the total amount that leaks from joint $i$ is $\delta_i^{w}$ ($= c_i^{w} \bar q_i$). This is illustrated by (the bold arrow in) Figure \ref{fig_decision_share}.

		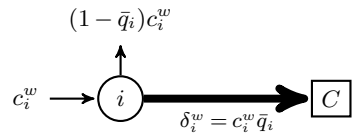
\begin{figure}[!h]	
			\begin{center}
			\caption{Decision Share: The View from Agent $i$}
			\label{fig_decision_share}
			\begin{tikzpicture}[->,>=stealth',shorten >=1pt,auto,node distance=3cm,thick,
				agent node/.style={circle,draw,font=\sffamily\bfseries\footnotesize},
				choice node/.style={rectangle,draw,font=\sffamily\footnotesize},
				invis node/.style={font=\sffamily\footnotesize},
				every node/.style={font=\sffamily\scriptsize},
				x=0.70cm,y=0.70cm
				]
			
			  \node[agent node] (i) at (0,0) {$i$};
			  \node[invis node] (NetOut) at (0,1.5) {$(1-\bar q_i) c_i^{w}$};
			  \node[invis node] (NetIn) at (-1.8,0) {$c_i^{w}$};
			  \node[choice node] (C) at (4,0) {$C$};
			  \node[invis node] (C Balance) at (-4,0) {};
			
			  \path[every node]
			    (NetIn)	edge node [below] {} (i)
			    (i)		edge [line width=3pt] node [below] {$\delta_i^{w}=c_i^{w}\bar q_i$} (C)
						edge node [right] {} (NetOut);
			\end{tikzpicture}
			\end{center}
		\end{figure}

%
%
}


\subsection{Generalization: Beyond Finite Choice Sets.}\label{ss_decision_share_generalization}

	Decision share provides a convenient route to generalization. Suppose $C$ were no longer a finite set. Taking the place of the $q^{(j)}$'s would be a collection of probability measures on $C$, one for each agent (the set of agents remaining finite).
In this setting the decision shares remain well-defined, being fully determined by $w$ and $P$.
Qualitatively, the resulting ``choice share distribution'' of the total endowment $\sum_{i\in A}{w_i}$ would be a ``mixture'' of the aforementioned probability distributions (over $C$) associated with the various agents, with the decision shares as the ``mixing weights''.

Specifically, suppose each agent $i\in A$ had preferences over $C$ described by probability measure $\mu_i$. Consider a subset $S \subseteq C$. The choice share of $S$ is then
\begin{equation}\label{decision_share_generalization}
	\sum_{i\in A} {\delta_i^{w} \mu_i(S)},
\end{equation}
which is natural and intuitive given the definition of decision share.

Equation (\ref{decision_share_generalization}) suggests that one may view choice outcomes as arising from a latent class model with $|A|$ classes of choice behaviors described by $\{\mu_i\}_{i\in A}$. In particular, an agent with index $i$ (agent $i$) behaves like ``someone'' in class $k$ with probability $\delta_k^{e_i}$. Alternatively, $\{\delta_k^{e_i}\}_{k\in A}$ describes the impact of the various agents in $A$ on agent $i$'s eventual choice.


\subsection{A Comparison with \cite{Kempe2003,Kempe2005}.}\label{ss_comparison_KKT}

	It would be useful to touch on how the model relates to those presented in \cite{Kempe2003,Kempe2005}. In particular, it is shown that this choice model is not just a special case of their models with multiple ``activation'' categories.
To this end, following an equivalence result of \cite{Kempe2005}, it would suffice to consider their general ``decreasing cascade model''.

	Consider using Monte Carlo simulation of some ``cascade process'' constructed to replicate the expected outcome of the model.
Sample runs should be based on equation (\ref{partition_of_activity}), which outlines the possible actions of each agent, and equations (\ref{choice_model_0})-(\ref{choice_model_2}) should be deducible from the stipulated dynamics. Thus leading, under Assumption \ref{assumption}, to equation (\ref{choice_share_eq}).

	This may be sketched out as follows, begin by sampling the agents who select a choice without consulting their neighbors (and their choices) based on $\{q^{(j)}\}_{j\in C}$. (For alignment with the ``decreasing cascade model'', let some choice $u\in C$ correspond to ``non-activation''.)
A kind of ``cascade process'' will be used to determine the choices of the remaining agents, who will adopt the choices of others.
Following \cite{Kempe2003,Kempe2005}, let agents who have already made choices in $C$ (less $u$) be known as ``activated'' agents. For each ``unactivated'' (undecided) agent $i$ except those that picked $u$ initially, each ``activated'' agent $k$ with $p_{ik}>0$ will have, at most, one opportunity to have the former agent adopt the the ``activated'' agent's choice (with some probability to be described later), thus ``activating'' the undecided agent (none if $p_{ik}=0$). Once no ``activation attempts'' remain, the process terminates.
Naturally, instances where (i) not all agents make a choice, or where (ii) no agent picks $u$ initially but some agent remains unactivated, will be discarded.

	In line with equation (\ref{partition_of_activity}), the aforementioned ``activation probabilities'' may be computed from $\{p_{ik}\}_{i\in A, k\in A}$ by conditioning on the set of agents whose choices were not adopted (``failed activation attempts'') and also on the event that an ``unactivated'' agent did not make a choice without consulting his/her network (beginning ``unactivated''). 
So, one may indeed deduce equations (\ref{choice_model_0})-(\ref{choice_model_2}) from the stipulated dynamics.
This is consistent with the interpretation that each agent selects from an ``extended choice set'' comprising choices in $C$ and the adoption of the choices of other agents. 
The sequencing of ``activation attempts'' may be done arbitrarily as, equivalently, an outcome is fully determined by a sample (with rejection) from a discrete distribution with $|A|+|C|-1$ categories for each agent $i\in A$ with selection probabilities $\{p_{ik}\}_{k\in A, k\not=i}$ and $\{q_{ij}\}_{j\in C}$.

	In the above ``cascade process'', the activation probabilities are, for each undecided agent, strictly increasing in the set of agents who had previously ``attempted'' and failed to ``activate'' the undecided agent. That violates the assumptions of the ``decreasing cascade model'' which requires those probabilities to be decreasing.
This is because each ``failed activation attempt'' increases the set of actions (to adopt the choice of some other agent) that were not taken. Conditional on the aforementioned set of actions not being taken, the probability of each of the remaining actions necessarily increases in that set of actions. Example \ref{ex_increasing_cascade} will make this clear.

\begin{example}[An ``Increasing'' Cascade]\label{ex_increasing_cascade}
		Consider, again, Example \ref{ex_influential_agent} and an outcome defined by the bold arrows of Figure \ref{fig_ex_increasing_cascade}. Agents 1 and 2 have both selected choice $B$ without consulting others in the network (let $A$ correspond to $u$). They then will attempt to ``activate'' agent 3 (get agent 3 to adopt their choices) according to the ``cascade process'' sketched out above.

		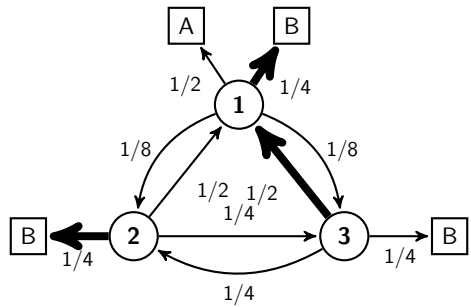
\begin{figure}[!h]	
			\begin{center}
			\caption{A visual representation of Example \ref{ex_increasing_cascade}.}
			\label{fig_ex_increasing_cascade}
			\begin{tikzpicture}[->,>=stealth',shorten >=1pt,auto,node distance=3cm,thick,
				agent node/.style={circle,draw,font=\sffamily\bfseries\footnotesize},
				choice node/.style={rectangle,draw,font=\sffamily\footnotesize},
				every node/.style={font=\sffamily\scriptsize},
				x=0.70cm,y=0.70cm
				]
			
			  \node[agent node] (1) at (0,0) {1};
			  \node[choice node] (1A) at (-1,1.5) {A};
			  \node[choice node] (1B) at (1,1.5) {B};
			  \node[agent node] (2) at (-2,-2.5) {2};
			  \node[choice node] (2B) at (-4,-2.5) {B};
			  \node[agent node] (3) at (2,-2.5) {3};
			  \node[choice node] (3B) at (4,-2.5) {B};
			
			  \path[every node]
			    (1) edge [bend left] node[right] {1/8} (3)
			        edge [bend right] node[left] {1/8} (2)		        
			        edge node[below left] {1/2} (1A)		        
			        edge [line width=3pt] node[below right] {1/4} (1B)		        
			    (2) edge node [below right] {1/2} (1)
			        edge node {1/4} (3)
			        edge [line width=3pt] node [below, line width=3pt] {1/4} (2B)		        
			    (3) edge [line width=3pt] node [below left] {1/2} (1)
			        edge [bend left] node {1/4} (2)
			        edge node [below] {1/4} (3B);		        
			\end{tikzpicture}
			\end{center}
		\end{figure}
		If agent 1 attempted to ``activate'' agent 3 first, the probability of ``activation'' would be $\frac{p_{31}}{p_{31}+p_{32}}=\frac{2}{3}$, and if agent 2 attempted to ``activate'' agent 3 after agent 1 failed, ``activation'' would occur with certainty ($\frac{p_{32}}{p_{32}} = 1$). Similarly, if agent 2 attempted to ``activate'' agent 3 first, the probability of ``activation'' would be $\frac{p_{32}}{p_{31}+p_{32}}=\frac{1}{3}$, and if agent 1 attempted to ``activate'' agent 3 after agent 2 failed, ``activation'' would occur with certainty ($\frac{p_{31}}{p_{31}} = 1$). So for a given ``activated'' agent, ``activation'' probabilities are strictly increasing in the set of agents who failed to activate agent 3. \hfill$\square$
\end{example}

	Thus, general as the models of \cite{Kempe2003,Kempe2005} are, the model presented here is not a mere special case featuring multiple ``activation'' categories.


\section{Selected Applications of the Model}

	Now, having previously touched upon the descriptive aspects of the model, to illustrate some of the possible ways to put the model to work, some applications for prescription will be touched upon. Specifically, an application of the model to identifying ``influential'' consumers for social marketing will be presented, followed by a more traditional example on pricing and price competition.


\subsection{Social Marketing: Brand Ambassador Selection}
\label{ss_influence_max}

	In the introduction, a great deal was said about social marketing. 
In this sub-section, a problem faced by the many marketing departments is cast in terms of this model. Specifically, as businesses seek to co-opt consumers as product evangelists, they face the problem of identifying who to reach out to so as to maximize the effectiveness of their marketing dollar.

	Here, the ``brand ambassador'' selection problem will be introduced, and it will be shown that though it is NP-hard, it admits and efficiently computable approximate solution that is guaranteed to be at least $1-1/e$ (about 63\%) as good as the optimal solution. 
	
	Given some choice $j\in C$ (that represents the brand whose choice share one seeks to maximize), the objective of the ``brand ambassador'' selection problem is to pick a subset $B\subseteq A$ of at most $K$ agents, who will select choice $j$ exclusively, such that the choice share of $j$ is maximized. Making $P$ and $\{q^{(l)}\}_{l\in C}$ functions of the set of brand ambassadors selected, brand ambassadors' exclusive recommendation of $j$ may be modeled as follows:

With apologies for the abuse of notation, for $B\subseteq A$, let
\begin{equation}
	[P(B)]_{ik} = \left\{
			\begin{array}{ll}
				0 & (i\in B) \\
				p_{ik} & (i\not\in B)
			\end{array}
		\right.
\end{equation}
and
\begin{equation}
	[q^{(l)}(B)]_{k} = \left\{
			\begin{array}{ll}
				0 & (k\in B, l\not=j) \\
				1 & (k\in B, l=j) \\
				q_{kl} & (k\not\in B).
			\end{array}
		\right.
\end{equation}
These mean that agents selected as brand ambassadors are modeled as changing their behavior such that they recommend and select choice $j$ only (with probability $1$). Other agents are unaffected.

	(As a technical digression, note that if Assumption \ref{assumption} is valid when $B = \emptyset$, then it is valid for all $B\subseteq A$. This is because $P = P(\emptyset)\ge P(B)$ entry-wise for all $B\subseteq A$, implying that the spectral radius of $P(\emptyset)$ is greater than $P(B)$ since both are non-negative matrices.)

	The brand ambassador selection problem may be expressed as follows:
	\begin{equation}
		\label{BA_problem}
		\max\{\pi_j^{w}(B) : B \subseteq A, |B| \le K \}.
	\end{equation}

	Due to its combinatorial nature, it is not surprising that:
	\begin{proposition}
		\label{BA_NP_hard}
		The optimization problem (\ref{BA_problem}) is NP-hard.
	\end{proposition}

	In spite of this result, it remains possible to obtain good solutions for the brand ambassador selection problem.
	It will be demonstrated that problem (\ref{BA_problem}) admits an efficient $(1-\frac{1}{e})$ approximation via a greedy selection strategy that may be described as follows:
	\begin{algorithm}[H]
	\caption{Greedy Algorithm (subject to maximum cardinality $K$)}
	\label{greedy_alg}
		\begin{algorithmic}
		\State $S_0 \gets \emptyset$

		\For{$i = 1$ to $K$} 
			\State $a_i^* \gets \arg\max\limits_{a_i\in A\backslash S_{i-1}} \pi_j^{w}(S_{i-1} \cup \{a_i\})$
			\State $S_i \gets S_{i-1} \cup \{a_i^*\}$
		\EndFor

		\State \Return $S_K$
		\end{algorithmic}
	\end{algorithm}

	As in \cite{Kempe2003,Kempe2005}, the proof of approximability makes use of a classic result due to \cite{Nemhauser1978} on the approximate maximization of monotone submodular functions with the greedy algorithm.
	A submodular function $f$ over a set $\Omega$ is a set function such that for every $X, Y \subseteq \Omega$ with $X \subseteq Y$ and every $x \in \Omega \backslash Y$, $f(X\cup \{x\})-f(X)\ge f(Y\cup \{x\})-f(Y)$ (see Chapter 44 of \cite{Schrijver2004}).
	A monotone submodular function is a submodular function such that for every $X, Y \subseteq \Omega$ with $X \subseteq Y$, $f(Y)\ge f(X)$.
	(Notably, the submodularity proof in \cite{Kempe2005} relies on the ``activation probabilities'' possessing the decreasing property. This is not present in the natural ``cascade model'' that generates the results of this model (\textsection\ref{ss_comparison_KKT}).)

	\begin{theorem}[\cite{Nemhauser1978}]\label{nemhauser_et_al_greedy_max}
		If the greedy algorithm is used for the approximate maximization of a non-negative monotone submodular set function $f$ over a set $\Omega$, the value of the solution generated by the greedy algorithm when it terminates at a set of size $K$, $S_K^{\rm Greedy}$, satisfies
	$$f(S_K^{\rm Greedy}) \ge \left(1-\frac{1}{e}\right) \max_{S\subseteq\Omega\atop |S|\le K} {f(S)}.$$
	\end{theorem}
	(The interested reader may refer to \cite{Krause2012} for more details on the maximization of monotone submodular functions, including a minor extension to the above result.)

	In the brand ambassador selection problem, the incremental choice share due to adding an additional brand ambassador is always non-negative (choice share is monotone in $B$):
	\begin{lemma}[Incremental Benefit of an Additional Brand Ambassador]
		\label{effect_new_BA}
		If $a\in A\backslash B$,
		\begin{equation}
			\label{extra_BA}
			\pi_j^{w}(B\cup\{a\}) - \pi_j^{w}(B) = w^{T}M_B e_a\frac{\sum_{l\in C\backslash\{j\}}{q_{al}} + \sum_{k\in A}{p_{ak}\left(1-\pi_{kj}(B)\right)}}{1 + p_a^{T}M_B e_a} \ge 0,
		\end{equation}
		where $M_B = (I-P(B))^{-1}$ and $p_a^{T}$ is the row of $P$ corresponding to agent $a$.
	\end{lemma}

	Before going on to establish the submodularity of $\pi_j^{w}(\cdot)$, it would be instructive to discuss equation (\ref{extra_BA}) in the context of some recent empirical findings.
	In a recent study, \cite{Godes2009} used data from a large-scale field test and an online experiment to study the efficacy of the proactive management of customer-to-customer communication by firms.
Among other findings, they provided evidence that it is not necessarily ``highly loyal'' customers who generate the important incremental ``word-of-mouth'' (WOM) as one might have expected, and that it may be more beneficial for firms to target ``less loyal'' customers.

	In equation (\ref{extra_BA}), $w^{T}M_B e_a$ may be viewed as a weighted centrality measure for agent $a$. (Note that $M_B$ decreases entry-wise in $B$, so the centrality of each agent decreases as more brand ambassadors are added.) The larger $w^{T}M_B e_a$ is, the larger the choice share contribution from including agent $a$. This accords well with the ``folk'' practice, mentioned in \cite{Godes2009}, wherein firms seeking to engineer WOM begin by attempting to identify ``key influencers''.

	On the other hand, including agent $a$ would not accord $j$ an incremental benefit amounting to all of agent $a$'s ``residual'' centrality (given the current set of brand ambassadors). Firstly, agent $a$ may already been selecting/recommending choice $j$ himself/herself. Furthermore, $a$'s inclusion causes an indirect reduction in the choice share of $j$ because agent $a$ would have, prior to being included in the set of brand ambassadors, selecting choice $j$ due to adopting the choices of other agents. This is why, as suggested by \cite{Godes2009}, firms seeking to engineer WOM should identify ``key influencers'' outside ``communities'' of ``loyal users''.

	While the equation (\ref{extra_BA}) hints at the submodularity of $\pi_j^w(\cdot)$, it is insufficient to establish submodularity as $M_B$ decreases entry-wise in $B$. Still, submodularity can indeed be demonstrated:

	\begin{proposition}[Submodularity in the Brand Ambassador Setting]
		\label{BA_submodularity}
		$\pi_j^{w}:A\rightarrow \Real$ is monotone and submodular for all $w\ge 0$.
	\end{proposition}

	With this, it follows immediately from Theorem \ref{nemhauser_et_al_greedy_max} that:	

	\begin{theorem}[Approximability in the Brand Ambassador Setting]\label{ba_problem_approximable}
		The optimization problem (\ref{BA_problem}) admits a $\left(1 - \frac{1}{e}\right)$-optimal greedy approximation described by Algorithm \ref{greedy_alg}.
	\end{theorem}

	When agents are associated with heterogeneous costs, one has to deal with more general knapsack constraints. \cite{Sviridenko2004} developed a variant of the greedy algorithm that produces a $\left(1 - \frac{1}{e}\right)$ approximation, but the computational cost is considerably higher ($O(n^5)$ function evaluations).
	Further discussion on this application is deferred to Appendix \ref{appendix_more_BA} of the online supplement.


\subsection{Exogenous Parameters and Choice Share: Optimization}
\label{ss_optimization}

	One of the most important business applications of discrete choice is the study of the variation of choice probabilities with respect to (exogenous) changes in problem parameters such as price and the ``design'' of choices (for instance, product characteristics). Once estimated from data, choice models are applied in the pricing of offerings or even in product-line design.

Early in this paper, the model presented was cited as a ``meta-model'' of choice, which might have attracted some protest at the time because $\{q^{(j)}\}_{j\in C}$ was constant, thus apparently precluding the aforementioned applications. In conjunction with the foregoing discussion of the brand ambassador selection problem, this sub-section will make clear that no such restriction exists.

Here, the model will be generalized to relate the choice share to an exogenous parameter. This will be illustrated, mainly, by way of an extended example on pricing/price competition.


\subsubsection{Extending the Model.}
\label{sss_extending_model}

	Suppose that for all $i\in A$, $k\in A$, $l\in C$, $p_{ik}(\cdot)$ and $q_{il}(\cdot)$ are functions of a (possibly multi-variate) parameter $u\in\Rn{d}$ with base values $P(0)=P$ and $q^{(j)}(0)=q^{(j)}$ for all $j\in C$. Once again, with apologies for the abuse of notation, define $P(u)$ to be the matrix containing the entries $\{p_{ik}(u)\}_{i\in A, k\in A}$, $q^{(j)}(u)$ the vector containing the entries $\{q_{ij}(u)\}_{i\in A}$, and
	\begin{equation} \label{choice_model_w_eq_param}
		\pi_{j}^{w}(u) = w^{T}(I - P(u))^{-1}q^{(j)}(u)
	\end{equation}
	to be the choice share of $j$ given endowment $w$ ($w\ge 0$) and parameter $u$.
	Naturally, a mild assumption will be made to ensure a meaningful mapping from $u$ to the $p_{ik}(\cdot)$'s and the $q_{ij}(\cdot)$'s:

	\begin{assumption}\label{assumption2}
		There exists a non-empty set $T \subseteq \Rn{d}$ such that for all $u\in T$, Assumption \ref{assumption} holds for $P(u)$ and $\{q^{(j)}(u)\}_{j\in C}$, $P(u)$ is a non-negative matrix, $q^{(j)}(u)$ is non-negative for all $j\in C$, and $\sum_{j\in C}{q_{ij}(u)} + \sum_{k\in A}{p_{ik}(u)} = 1$ for all $i\in A$. Furthermore, without loss of generality, $0\in T$ and $P(0)=P$ and $q^{(j)}(0)=q^{(j)}$ for all $j\in C$.
	\end{assumption}

	Consider the following example illustrating how this framework may be applied:
	\begin{example}[Univariate Affine Variation of the Preferences of a Single Agent]\label{ex_linear_variation}
		Consider the special case of ``affine variation'' in the preferences of some agent $\bar r \in A$, where $P(u) = P - u e_{\bar r} v^{T}$ with
$q^{j}(u) = q^{j} + u e_{\bar r}$, $q^{l}(u) = q^{l} - u \beta_l e_{\bar r}$ for all $l\in C\backslash\{j\}$, and $\sum_{k\in A}{v_k} + \sum_{l\in C\backslash\{j\}}{\beta_{l}} = 1$. Then the following expression for the choice share of $j$:
		\begin{equation}
			\pi_{j}^{w}(u) = \pi_{j}^{w}(0) 
						 	+ u \theta_{i \bar r}(0)\frac{1 - \sum_{k\in A}{v_k \pi_{j}^{e_k}(0)}}{1 + u \sum_{k\in A}{v_k \theta_{k\bar r}(0)} } \label{linear_variation}
		\end{equation}
		holds on an interval containing $0$ where $\theta_{ik}(u)$ is defined in Lemma \ref{derivatives} of Appendix \ref{appendix_pfs}.

		In the case of ``pure preference shifting'' where $v=0$, equation (\ref{linear_variation}) reduces to $\pi_{j}^{w}(u) = \pi_{j}^{w}(0) + u \theta_{i \bar r}(0)$.
		Equation (\ref{linear_variation}) may be obtained using the same method used in Lemma \ref{effect_new_BA}.
		Furthermore, Lemma \ref{effect_new_BA} can be easily verified to be a special case of equation (\ref{linear_variation}).
		\hfill$\square$
	\end{example}


\subsubsection{Pricing and Price Competition.}

	As an extended example, a simple model of pricing/price competition will be presented. Here, some strict subset of choices $D\subset C$ will correspond to products sold by different firms which are engaged in price competition. (It may be taken that $C\backslash D$ contains an ``outside option''.) Here, endowments should be interpreted as ``units of choice'', and choices and firms will be referred to interchangeably using elements of $D$.

	For $j\in D$, denote the base profit for $j$ per ``unit of choice'' by $m_j$ ($m_j>0$), and denote the firm's (real-valued) decision variable, the price discount to be granted, as $z_j$. Denote the vector of price discounts for $D\backslash\{j\}$ as $z_{-j}$.
Of course, the model parameters are determined by $\{z_l\}_{l\in D}$.

	Suppose that, for $j\in D$, the parameters may be described as univariate functions for any given $z_{-j}$.
The following model that describes substitution into a choice as it becomes more attractive:
Given $z_{-j}$, suppose that the set $T$ on which Assumption \ref{assumption2} is a non-empty interval and on $T$,
	\begin{itemize}
		\item \label{pricing_qij_strictly_monotone}
			$q_{ij}(\cdot, z_{-j})$ is concave and strictly increasing, and
		\item
			$q_{il}(\cdot, z_{-j})$ ($l\in C$, $l\not=j$) and $p_{ik}(\cdot, z_{-j})$ ($k\in A$), are convex and decreasing.
	\end{itemize}
	(Appropriately defined affine functions satisfy this.)
	Suppose, in addition, that for $j\in D$, firm $j$ is limited to offering price discounts $z_j\in S_j := [L_j, U_j]$ where and $L_j > -\infty$ and $U_j \le m_j$.


	For $(z_j,z_{-j})$ where choice shares are well-defined, $\pi_j^w(z_j,z_{-j})$ may be defined analogously as before and the profit for $j\in D$ denoted as
	\begin{equation}
		\Pi_j(z_j,z_{-j}) = (m_j - z_j) \pi_{j}^{w}(z_j,z_{-j}).
	\end{equation}
	Otherwise, let the profit for $j\in D$ be $-\infty$.
Thus, given $z_{-j}$, firm $j$'s profit is concave in $z_j$:
	\begin{proposition}[Concavity of Profit]\label{concavity_profit}
		For all $j\in D$, $\Pi_j(\cdot, z_{-j})$ is concave on $S_j$.
	\end{proposition}
	This means that the associated single player pricing problem may be efficiently solved.
Furthermore, Proposition \ref{concavity_profit} allows one to quickly deduce that:
	\begin{theorem}[Equilibrium in Pricing Game] \label{equilibrium_in_pricing_game}
		The game $G=(\{\Pi_j\}_{j\in D}, \{S_j\}_{j\in D})$ has a pure strategy Nash equilibrium.
	\end{theorem}


\section{Concluding Remarks}

	In this paper, a simple model of choice where decisions are influenced by peers on a recommendation network was introduced premised on the notion that the presence of a recommendation network extends the set of available decisions to include choice adoption behavior. An immediate outcome of the modeling assumptions is that each agent's choice probabilities turns out to be a kind of ``mixture'' of the ``personal choice probabilities'' of all agents in the network.

\BlueText{
	Through efficiently computable closed form solutions, the model readily yields insights on how agents in a recommendation network make choices. In particular, the concept of decision share, introduced in this paper, neatly shows how ``influence'' in a recommendation network is driven by both ``centrality'' and ``decisiveness''. The natural manner through which decision share enables generalization to more generic choice sets suggests that it is a natural measure of ``influence'' in a recommendation network.
}

	The model is potentially applicable to large-scale influence maximization computations. In particular, the brand ambassador selection problem was introduced and shown to admit efficiently computable $(1-\frac{1}{e})$-optimal approximations via a greedy selection strategy. A generalization to continuous choice sets was also described through the use of the notion of ``decision share'', a natural adjoint to ``choice share'', allowing the model to be used in more general settings.

	Finally, it bears repeating that the model presented in this paper may be considered a ``meta-model'' of choice. In principle, any choice model may be extended to include adopting the action of another agent as an extension of the choice set, allowing it to be ``plugged in'' to this model, thus enabling one to extend existing choice models to model social network effects.



%
%
%

\vspace{3ex}
\begin{APPENDICES}
\section{Proofs}\label{appendix_pfs}


	\proof{Proof of Lemma \ref{M_well_defined_etc}.}
		If (\ref{spectral_radius_property}) holds, then the sum $I+P+P^2+\ldots$ is well-defined, implying (\ref{M_well_defined_property}). Conversely, (\ref{M_well_defined_property}) cannot hold unless (\ref{spectral_radius_property}) holds. So (\ref{spectral_radius_property}) and (\ref{M_well_defined_property}) are equivalent.
		Clearly, (\ref{subM_well_defined_property}) implies (\ref{M_well_defined_property}).
To see how (\ref{M_well_defined_property}) implies (\ref{subM_well_defined_property}), consider the matrix $P_V$ formed by keeping the entries corresponding to $V$ and setting the rest to $0$. For $n\ge 1$, the entries of $P_V^n$ and the corresponding entries of $V^n$ are equal and $P^n$ dominates $P_V^n$ entry-wise. Therefore, $(I-P_V)^{-1}$ is well-defined and so is $(I-V)^{-1}$. So (\ref{spectral_radius_property}), (\ref{M_well_defined_property}) and (\ref{subM_well_defined_property}) are equivalent.

		To show that Assumption \ref{assumption} implies (\ref{spectral_radius_property}), first note that the Perron-Fr\"obenius Theorem states that $P$, being a non-negative matrix has a positive real eigenvalue $\lambda$ that is maximal in terms of magnitude with corresponding left eigenvector, $v$, having strictly positive terms. Without loss of generality, suppose $v^{T}v = 1$ so $v \le e$. Let $\epsilon = \left(\min_{i\in Q}{\sum_{j\in C}{q_{ij}}} \right) \left(\min_{i\in A\backslash Q}{\prod_{k=1}^{n_i}{p_{n_k^i n_{k+1}^i}}}\right)$ ($\epsilon > 0$ by Assumption \ref{assumption}), and let $N = \max_{i\in A\backslash Q}{n_i}$.
Since the row sums of $P$ are non-negative but at most unity, for all $m,n>0$ and vectors $u\ge 0$, $u^{T}P^n e \ge u^{T}P^{n+m} e \ge 0$.
Therefore, for all $m>0$, $\lambda^{mN} = v^{T}P^{mN} v \le v^{T}P^{mN} e \le \sum_{i\in A} v_i e_i^{T}P^{m n_i} e \le (1-\epsilon)^m v^{T}e$. This implies that $\lambda < 1$, meaning (\ref{spectral_radius_property}) holds.

		To show how (\ref{spectral_radius_property}) implies Assumption \ref{assumption}, the contrapositive will be proven. Suppose Assumption \ref{assumption} does not hold. If $Q=\emptyset$ then $P$ is a row-stochastic matrix and $1$ is an eigenvalue of $P$ with its corresponding left eigenvector consisting of all ones, so suppose $Q\not=\emptyset$. Let $Y$ be the subset of $A\backslash Q$ for which no path satisfying the requirements of Assumption \ref{assumption} exists. For all $i\in Y$, by equation (\ref{partition_of_activity}), $\sum_{k\in Y}{p_{ik}}=1$. Otherwise a path satisfying the requirements of Assumption \ref{assumption} may be constructed by augmenting a path from an agent in $(A\backslash Q)\backslash Y$ to $Q$. Thus, the sub-matrix of $P$ corresponding to the agents in $Y$ is row-stochastic. So $1$ is an eigenvalue of $P$ with its corresponding left eigenvector having $1$'s in the entries corresponding to agents in the set $Y$ and $0$'s elsewhere. So Assumption \ref{assumption} implies (\ref{spectral_radius_property}), meaning Assumption \ref{assumption} and (\ref{spectral_radius_property}) are equivalent, completing the proof.
\hfill$\blacksquare$
	\endproof


	\proof{Proof of Proposition \ref{BA_NP_hard}.}
		The result will be proven via a reduction of the Vertex Cover Problem which is known to be NP-complete. Consider an instance of the Vertex Cover Problem defined by an undirected graph $G=(V,E)$ and an integer $K$. Therein, a set, $S$, of $K$ vertices is sought such that every edge in $E$ has at least one end point in $S$.

		Create an instance of the Brand Ambassador Selection Problem with $A=V$ and $C=\{\alpha,\beta\}$, with $j=\alpha$ and $w$ as the vector of all ones. For all $i\in A$, $q_{i\alpha}=0$ and $q_{i\beta}=1/2$. (Note that Assumption \ref{assumption} is satisfied.) Denote $\eta(i)$ as the number of vertices adjacent to $i$ in the graph $G$ and let $\overline{\eta} = \max\limits_{i\in A}{\eta(i)}$. For each $i\in A$, let $p_{ik}=0$ if $(i,k)\not\in E$ and $p_{ik}=\frac{1}{2\eta(i)}$ otherwise.

		For any vertex cover $S$, $P(S)$ and $\{q^{(l)}(S)\}_{l\in C}$ are such that agents in $A\backslash S$ never directly adopt each other's choices. It then follows that $\pi_{\alpha}^{w}(S) = \frac{1}{2}(n+K)$.
		But for any subset of A, $\tilde S$ where $|\tilde S| = K$ and $\tilde S$ is not a vertex cover. Then there exists $(r,s)\in E$ such that $r,s\not\in \tilde S$. This implies that $\pi_{\alpha}^{w}(\tilde S) \le \frac{1}{2}(n+K) - \frac{1}{4\overline{\eta}} < \pi_{\alpha}^{w}(S)$.

		By Lemma \ref{effect_new_BA}, $\pi_{\alpha}^{w}(\cdot)$ is monotone. Therefore any optimal solution to this instance of the Brand Ambassador Problem may be (trivially) identified with a solution to the Vertex Cover Problem. In fact, a bijection exists between the set of optimal solutions for brand ambassador problem and the set of vertex covers. Therefore, the Brand Ambassador Problem is at least as computationally hard as the Vertex Cover Problem.
		\hfill$\blacksquare$
	\endproof


	\proof{Proof of Lemma \ref{effect_new_BA}.}
		By the Sherman-Morrison-Woodbury formula 
for an invertible matrix $A$, if $A+uv^{T}$ is invertible, then
		\begin{equation}
			(A+uv^{T})^{-1} = A^{-1} - \frac{A^{-1}uv^{T}A^{-1}}{1+v^{T}A^{-1}u}.
		\end{equation}
		The corollary follows by applying this to
		$$
			\pi_j^{w}(B\cup\{a\}) = w^{T}(I - P(B) + e_a p_a^{T})^{-1}\left(q^{(j)}(B) + \left(p_a^{T}e + \sum_{l\in C\backslash\{j\}}{q_{al}}\right)e_a\right),
		$$
		and rearranging terms. (In doing so, note that $p_a^{T}M_Bq^{(j)}(B) = \sum_{k\in A} {p_{ak} \pi_{kj}(B)}$.)
		\hfill$\blacksquare$
	\endproof


	\proof{Proof of Proposition \ref{BA_submodularity}.}
		Monotonicity follows from Lemma \ref{effect_new_BA}. To prove submodularity, it suffices to verify that for all $X \subseteq Y \subseteq A$ and $a\in A\backslash Y$,
		$$
			\pi_j^{w}(X\cup\{a\}) - \pi_j^{w}(X) \ge \pi_j^{w}(Y\cup\{a\}) - \pi_j^{w}(Y)
		$$
		and to do so for all non-negative $w$. Thus, it is necessary and sufficient to show that for all $i \in A$,
		\begin{equation} \label{submodularity_relation}
			\pi_j^{e_i}(X\cup\{a\}) - \pi_j^{e_i}(X) \ge \pi_j^{e_i}(Y\cup\{a\}) - \pi_j^{e_i}(Y)
		\end{equation}
		for arbitrary $X$, $Y$, $a$ satisfying $X \subseteq Y \subseteq A$ and $a\in A\backslash Y$.

		Before proceeding, first, observe that
		\begin{equation} \label{brand_ambassador_only_choice}
			\pi_j^{e_i}(X) = 1\ {\rm if}\ i\in B.
		\end{equation}

		Now, consider the four possible cases, $i = a$, $i \in X$, $i \in Y\backslash X$ and $i \in A\backslash (Y \cup \{a\})$.
		
		When $i = a$, $\pi_j^{e_i}(X\cup\{a\}) - \pi_j^{e_i}(X) = 1 - \pi_j^{e_i}(X)$ and $\pi_j^{e_i}(Y\cup\{a\}) - \pi_j^{e_i}(Y) = 1 - \pi_j^{e_i}(Y)$ by equation (\ref{brand_ambassador_only_choice}). Since $\pi_j^{e_i}(Y) \ge \pi_j^{e_i}(X)$ by monotonicity, inequality (\ref{submodularity_relation}) holds.

		When $i \in X$, $\pi_j^{e_i}(X\cup\{a\}) = \pi_j^{e_i}(X) = \pi_j^{e_i}(Y\cup\{a\}) = \pi_j^{e_i}(Y) = 1$ by equation (\ref{brand_ambassador_only_choice}), so inequality (\ref{submodularity_relation}) holds.

		When $i \in Y\backslash X$, $\pi_j^{e_i}(X\cup\{a\}) - \pi_j^{e_i}(X) \ge 0$ by monotonicity and $\pi_j^{e_i}(Y\cup\{a\}) - \pi_j^{e_i}(Y) = 1 - 1 = 0$ by equation (\ref{brand_ambassador_only_choice}), so inequality (\ref{submodularity_relation}) holds.

		For each $i \in A\backslash (Y \cup \{a\})$, consider equations (\ref{choice_model_0})-(\ref{choice_model_2}) which defines each of $\pi_j^{e_i}(X\cup\{a\})$, $\pi_j^{e_i}(X)$, $\pi_j^{e_i}(Y\cup\{a\})$, and $\pi_j^{e_i}(Y)$. This may be written as
		\begin{equation}\label{submod_lin_sys_for_partial_inversion}
			\pi_j^{e_i}(S) - \sum_{k \not\in Y \cup \{a\}}{p_{ik} \pi_j^{e_k}(S)} =  q_{ij} + \sum_{k \in Y \cup \{a\}}{p_{ik} \pi_j^{e_k}(S)}
		\end{equation}
		for $S \cap (Y\cup\{a\}) = \emptyset$.
		For given $i \in A\backslash (Y \cup \{a\})$, the ``analogous'' coefficients and the constant remain the same for all such sets $S$. This is because the respective rows of $P$ and $q^{(j)}$ remain unchanged.

		By part (c) of Lemma \ref{M_well_defined_etc}, it follows that $\pi_j^{e_i}(S)$ may be expressed as a non-negative linear combination of $\{\pi_j^{e_k}(S)\}_{k \in Y \cup \{a\}}$ plus a non-negative constant:
		\begin{equation}\label{submod_lin_sys_partial_inverted}
			\pi_j^{e_i}(S) =  \rho_{ij}(S) + \sum_{k \in Y \cup \{a\}}{\gamma_{ik}(S) \pi_j^{e_k}(S)}.
		\end{equation}
		This is achieved by considering the linear system defined by (\ref{submod_lin_sys_for_partial_inversion}) for $i \in A\backslash (Y \cup \{a\})$ and ``pre-multiplying it'' by $(I-P_{A\backslash (Y \cup \{a\})})^{-1}$ where $P_{A\backslash (Y \cup \{a\})}$ is the principle sub-matrix of $P$ defined by the subset $A\backslash (Y \cup \{a\})$.

		For each $i\in A\backslash (Y\cup\{a\})$, the constants and the coefficients of $\{\pi_j^{e_k}(S)\}_{k \in Y \cup \{a\}}$ of equation (\ref{submod_lin_sys_partial_inverted}) are the same for every choice of $S\subseteq A$ satisfying $S \cap (Y\cup\{a\}) = \emptyset$.
Therefore, since inequality (\ref{submodularity_relation}) holds for all $i \in Y\cup\{a\}$, inequality (\ref{submodularity_relation}) holds when $i \in A\backslash (Y \cup \{a\})$.
		This completes the proof.
		\hfill$\blacksquare$
	\endproof


\begin{lemma}
	\label{derivatives}
	Suppose Assumption \ref{assumption} is valid for $P(u)$
	and let $\theta_{ik}(u) := e_i^{T} (I - P(u))^{-1} e_k$. For all $i\in A$,
	\begin{equation}
		\label{sensitivity_pi}
		\frac{d}{du}\pi_{j}^{e_i}(u)=
		\sum_{k\in A}{
			\theta_{ik}(u) G_{kj}(u)		
		},
	\end{equation}
	\begin{equation}
		\label{sensitivity_theta}
		\frac{d}{du}\theta_{ik}(u) =
		\sum_{s\in A}{
			\sum_{r\in A}{
				\theta_{ir}(u)\theta_{sk}(u) \frac{d}{du}p_{rs}(u)
			}
		}
	\end{equation}
	and
	\begin{equation}
		\label{second_deriv_pi}
		\frac{d^2}{du^2}\pi_{(i,j,C,P)}(u)=
		\sum_{k\in A}{
			\left[
				G_{kj}(u) \frac{d}{du}\theta_{(i,k,C,P)}(u)
				+
				H_{kj}(u) \theta_{(i,k,C,P)}(u)
			\right]
		}
	\end{equation}
	where
	\begin{equation} \label{def_G}
		G_{kj}(u) := \frac{d}{du}q_{kj}(u) + \sum_{s\in A}{\pi_{j}^{e_s}(u) \frac{d}{du} p_{ks}(u)}
	\end{equation}
	and
	\begin{equation} \label{def_H}
		H_{kj}(u) := \frac{d^2}{du^2}q_{kj}(u) + \sum_{s\in A}{ \pi_{j}^{e_s}(u) \frac{d^2}{du^2}p_{ks}(u)} + \sum_{s\in A}{\frac{d}{du} p_{ks}(u) \frac{d}{du}\pi_{j}^{e_s}(u)}.
	\end{equation}
\end{lemma}

	\proof{Proof.}
		Recall that for a differentiable univariate matrix function $M$, if $M(x)$ is non-singular,
		$$
			\frac{d}{dx} M^{-1}(x) = -M^{-1}(x)\left( \frac{d}{dx} M(x) \right) M^{-1}(x).
		$$
	By (\ref{spectral_radius_property}) of Lemma \ref{M_well_defined_etc}, $(I - P(u))^{-1}$ is well-defined, so through straightforward differentiation, one obtains
	$$
		\frac{d}{du}\pi_{j}^{e_i}(u)=
		e_i^{T} (I - P(u))^{-1}
		\left[
			\frac{d}{du} q^{(j)}(u)
			+
			\left( \frac{d}{du} P(u) \right)(I - P(u))^{-1} q^{(j)}(u)
		\right].
	$$
	By collecting terms, one arrives at equation (\ref{sensitivity_pi}). Equation (\ref{sensitivity_theta}) may be obtained in a similar fashion. Equation (\ref{second_deriv_pi}) may be obtained by differentiating equation (\ref{sensitivity_pi}).		
	\hfill$\blacksquare$
	\endproof


	\begin{lemma}[Monotonicity]\label{monotonicity_matrix_entries}
		For given $j\in C$, if for all $i\in A$, $q_{ij}(\cdot)$ is increasing, $p_{ik}(\cdot)$ are decreasing for all $k\in A$, and $q_{il}(\cdot)$ are decreasing for all $l\in C \backslash\{j\}$, then for a non-negative initial endowment $w$, $\pi_j^{w}(\cdot)$ is increasing on $T$, and for $l\in C\backslash\{j\}$, $\pi_l^{w}(\cdot)$ is decreasing on $T$.
	\end{lemma}	

	\proof{Proof.}
		The proof of this lemma makes use of Lemma \ref{derivatives} in Appendix \ref{appendix_pfs}.

		First, note that because $(I-P(u))^{-1} = I + P(u) + P(u)^2 + \ldots$ and $P(u)\ge 0$,
		\begin{equation} \label{theta_nonneg}
			\theta_{ik}(u) \ge 0
		\end{equation}
		for all $i,k\in A$.
	
		Now, for $k\in A$, $\sum_{s\in A}{p_{ks}(u)} + \sum_{l\in C}{q_{kl}(u)} = 1$, so
		\begin{align*}
			0	=\ & \frac{d}{du} q_{kj}(u) + \sum_{s\in A}{ \frac{d}{du} p_{ks}(u)}  + \sum_{l\in C\backslash\{j\}}{\frac{d}{du} q_{kl}(u)} \\
				\le\ & 	\frac{d}{du} q_{kj}(u) + \sum_{s\in A}{ \frac{d}{du} p_{ks}(u)} \\
				\le\ & 	\frac{d}{du} q_{kj}(u) + \sum_{s\in A}{ \pi_{j}^{e_s}(u) \frac{d}{du} p_{ks}(u) } \\
				=\ & G_{kj}(u)\ \ ({\rm defined\ by\ (\ref{def_G})\ in\ Lemma\ \ref{derivatives}})
		\end{align*}
		where the first inequality follows from the hypothesis that $q_{il}(\cdot)$ are decreasing for all $l\in C \backslash\{j\}$, and the second inequality follows from the fact that $\pi_{j}^{e_k}(u) \in [0,1]$ for all $k\in A$.
		Taken together with equation (\ref{sensitivity_pi}) of Lemma \ref{derivatives} and Corollary \ref{choice_share_w}, one may deduce that $\pi_j^{w}(\cdot)$ is increasing.

		For $l\in C\backslash\{j\}$, since $\pi_{l}^{e_s}(u) \ge 0$ for all $s\in A$ and, by hypothesis, the other terms on the right hand side of equation (\ref{def_G}) are non-positive, $G_{kl}(u) \le 0$. Thus, similarly, one may deduce that $\pi_l^{w}(\cdot)$ is decreasing.
	\hfill$\blacksquare$
	\endproof

	Notably, Lemma \ref{monotonicity_matrix_entries} generalizes Lemma \ref{effect_new_BA}.


	\begin{lemma}[Concavity/Convexity]\label{concavity_matrix_entries}
		Suppose the conditions of Lemma \ref{monotonicity_matrix_entries} hold and, in addition, $\forall\ i\in A$, $q_{ij}(\cdot)$ is concave, $p_{ik}(\cdot)$ is convex for all $k\in A$, and $q_{il}(\cdot)$ is convex for all $l\in C\backslash\{j\}$, then $\pi_j^{w}(\cdot)$ is concave on $T$, and for $l\in C\backslash\{j\}$, $\pi_l^{w}(\cdot)$ is convex on $T$.
	\end{lemma}

	\proof{Proof.}
		The proof of this lemma makes use of Lemma \ref{derivatives} in Appendix \ref{appendix_pfs}.

		As with (\ref{theta_nonneg}) in the proof of Lemma \ref{monotonicity_matrix_entries}, $\theta_{ik}(u) \ge 0$ for all $i,k\in A$. Since $p_{ik}(\cdot)$ is decreasing for all $i,k\in A$, in conjunction with equation (\ref{sensitivity_theta}), one obtains
		\begin{equation}
			\label{sign_d_theta}
			\frac{d}{du}\theta_{ik}(u) \le 0. 
		\end{equation}
		for all $i,k\in A$.
		
		In the proof of Lemma \ref{monotonicity_matrix_entries},
		it was previously shown that $G_{ij}(u)$ (see (\ref{def_G})\ in\ Lemma\ \ref{derivatives}) is non-negative.
		Based on equation (\ref{second_deriv_pi}) of Lemma \ref{derivatives}, 
		to show that $\pi_j^{e_i}(\cdot)$ is concave, it would be sufficient to show that $H_{kj}(u) \le 0$.
	
		For expositional convenience, equation (\ref{def_H}) is reproduced here:
		$$
			H_{kj}(u) := \frac{d^2}{du^2}q_{kj}(u) + \sum_{s\in A}{ \pi_{j}^{e_s}(u) \frac{d^2}{du^2}p_{ks}(u)} + \sum_{s\in A}{\frac{d}{du} p_{ks}(u) \frac{d}{du}\pi_{j}^{e_s}(u)}.
		$$
		
		The hypothesis that $p_{ik}(\cdot)$ is decreasing for all $i,k\in A$ and (\ref{sign_d_theta}) imply that
		$$
			\sum_{s\in A}{\frac{d}{du} p_{ks}(u) \frac{d}{du}\pi_{j}^{e_s}(u)} \le 0.
		$$
		
		Now, for $k\in A$ and any fixed $\gamma_s \in[0,1]$ for each $s\in A$, by manipulating equation (\ref{partition_of_activity}), one obtains
		$$
			q_{kj}(u) + \sum_{s\in A}{\gamma_s p_{ks}(u)} = 1 - \sum_{l\in C \backslash\{j\}}{q_{kl}(u)} - \sum_{s\in A }{(1-\gamma_s) p_{ks}(u)}.
		$$
		This implies that
		$$
			\frac{d^2}{du^2} q_{kj}(u) + \sum_{s\in A}{\gamma_s \frac{d^2}{du^2} p_{ks}(u)} = - \sum_{l\in C \backslash\{j\}}{\frac{d^2}{du^2} q_{kl}(u)} - \sum_{s\in A}{(1-\gamma_s) \frac{d^2}{du^2} p_{ks}(u)} \le 0
		$$
		where the inequality follows from the convexity of $q_{kl}(\cdot)$ for $l\in C \backslash\{j\}$ and the convexity of $p_{ks}(\cdot)$ for $s\in A$.
		
		Substituting $\pi_{j}^{e_s}(u)$ for $\gamma_s$, the fact that $H_{kj}(u)\le 0$ is established. Therefore, $\pi_j^{e_i}(\cdot)$ is concave. Combining this with Corollary \ref{choice_share_w}, it follows that $\pi_j^{w}(\cdot)$ is concave.
		
		Now consider $l\in C\backslash\{j\}$.
		In the proof of Lemma \ref{monotonicity_matrix_entries}, it has already been shown that $G_{il}(u) \le 0$. So it suffices to show that $H_{kl}(u) \ge 0$ for all $k\in A$. By Lemma \ref{monotonicity_matrix_entries}, $\frac{d}{du}\pi_{l}^{e_s}(u) \le 0$ for all $s\in A$. It then follows from the hypothesis of Lemma \ref{monotonicity_matrix_entries} that $H_{kl}(u) \ge 0$. In the same way as before, one may deduce that $\pi_l^{w}(\cdot)$ is convex.
	\hfill$\blacksquare$
	\endproof


	\proof{Proof of Proposition \ref{concavity_profit}.}
	By Lemma \ref{monotonicity_matrix_entries} and Lemma \ref{concavity_matrix_entries} of Appendix \ref{appendix_pfs}, the $\pi_{j}^{w}$'s are monotone and concave in the discounts granted. Therefore, where choice shares are well-defined,
	\begin{align}
		\frac{d^2}{dz_j^2}\Pi_j(z_j, z_{-j}) &\ = (m_j-z_j) \frac{d^2}{dz_j^2} \pi(z_j, z_{-j}) - 2\frac{d}{dz_j} \pi(z_j, z_{-j}) \\ &\ \le 0.
	\end{align}
	Noting that $\Pi_j(z_j,z_{-j}) = -\infty$ for $z_j$ where choice shares are not well-defined, it follows that $\Pi_j(\cdot, z_{-j})$ is concave on $S_j$.\hfill$\blacksquare$
	\endproof


	\proof{Proof of Theorem \ref{equilibrium_in_pricing_game}.}
	Since the pay-offs are concave on the respective $S_j$'s, which are compact convex sets, $G$ is a concave game. By a result due to Debreu, Glicksberg and Fan, $G$ has a pure strategy equilibrium (see, for instance, \textsection1.3.3 of \citealp{Fudenberg1991}).\hfill$\blacksquare$
	\endproof

\end{APPENDICES}



\ACKNOWLEDGMENT{The author would like to thank \textsc{Teo} Chung Piaw, Andrew \textsc{Lim} and Michael \textsc{Kim} for the time and energy they have contributed to critique the work and early drafts of the paper, and also for the many fun and fruitful discussions. 
}


\bibliographystyle{ormsv080} 
\bibliography{bib}		

\end{document}


\maketitle



%
%
%

\appendix

\vspace{-12ex}

\pagenumbering{arabic}
\renewcommand{\thepage}{S-\arabic{page}}
\setcounter{section}{1}
\titleformat{\section}{\large\bf}{Appendix \thesection:}{1em}{}
\renewcommand{\theequation}{\thesection\arabic{equation}}

\section{Parameter Estimation in a Restricted Setting}\label{appendix_estimation}

	Here, an approach for estimating $\{p_{ik}\}_{i\in A, k\in A}$ and $\{q^{(j)}\}_{j\in C}$ is described in a setting where empirical choice shares at an individual level ($\pi_{ij}$ for all $i\in A$, $j\in C$) are available, as well as information on the agents and their preferences, possibly via a questionnaire.

As in the work of \cite{Toubia2003}, the ``correct'' parameters are viewed to be an element of a polyhedral set, and information about the parameters may be used to reduce the size of the set.
In principle, any point within that set (or one within its relative interior) may be used as a point estimate.
(\cite{Toubia2003} focus on the analytic center as they use its properties to develop adaptive questionnaires for conjoint analysis.)

This Appendix discusses the formulation of a suitable set of linear constraints to define the set, after which an element within it may be drawn to obtain a point estimate of $\{p_{ik}\}_{i\in A, k\in A}$ and $\{q^{(j)}\}_{j\in C}$.

\subsection*{Fundamental Constraints}
	Equations (\ref{choice_model_0})-(\ref{choice_model_2}) describe the primary relationships that are to be respected: \begin{equation}
	\pi_{ij} = q_{ij} + \sum_{k \in A}{p_{ik} \pi_{kj}} \nonumber
\end{equation}
for all $i\in A$, $j\in C$.
However, in practice, it may be necessary to add ``slacks'' in case a perfect fit is not possible. This is because when data about the agents' preferences are used improve estimates (and to avoid the trivial solution $q_{ij}=\pi_{ij}$ for all $i\in A$, $j\in C$, and $p_{ik}=0$ for all $i,k\in A$), the problem may become infeasible due to possible inconsistency in that data. As such, one may consider, instead:
\begin{equation}\label{estimation_constraints_1}
	\pi_{ij} + \epsilon_{ij}^{+} - \epsilon_{ij}^{-} = q_{ij} + \sum_{k \in A}{p_{ik} \pi_{kj}}\ {\rm for\ all\ }i\in A,j\in C.
\end{equation}
Naturally, one would require that
\begin{equation}\label{estimation_constraints_2a}
	p_{ik}\ge 0\ {\rm for\ all\ }i,k\in A,
\end{equation}
\begin{equation}\label{estimation_constraints_2b}
	p_{ii} = 0\ {\rm for\ all\ }i\in A
\end{equation}
and
\begin{equation}\label{estimation_constraints_3}
	q_{ij}\ge 0, \epsilon_{ij}^{+}\ge 0, \epsilon_{ij}^{-}\ge 0 \ {\rm for\ all\ }i\in A,j\in C.
\end{equation}

In contrast to what was proposed for equations (\ref{choice_model_0})-(\ref{choice_model_2}), equation (\ref{partition_of_activity}) can be represented exactly:
\begin{equation}\label{estimation_constraints_4}
	\sum_{k \in A}{p_{ik}} + \sum_{j \in C}{q_{ij}}  = 1\ {\rm for\ all\ }i\in A.
\end{equation}

\subsection*{Incorporating Information}
	Subsequently, four classes of information on the agents will be modeled as constraints on agents' preferences:
(i) the relative importance of groups of agents,
(ii) relative preference between choices,
(iii) decisiveness of an agent, and
(iv) presence/absence of reliance on other agents (sparsity).

	\ \\
	\noindent{\em Relative importance of groups of agents.} A given agent $i\in A$ may rely more on a certain subsets of agents than another. This may be modeled as:
\begin{equation}\label{estimation_constraints_additional_info_1}
	\sum_{k_1 \in S_1}{p_{ik_2}} \ge K \sum_{k_2 \in S_2}{p_{ik_2}}
\end{equation}
where $S_1\subseteq A$, $S_2\subseteq A$ and $S_1\cap S_2 = \emptyset$ and for some $K$.

	\ \\
	\noindent{\em Relative preference between choices.} As above, a given agent may have known preferences across choices. This may be modeled as:
\begin{equation}\label{estimation_constraints_additional_info_2}
	(K-\Delta) q_{ij_2} \le q_{ij_1} \le (K+\Delta) q_{ij_2}
\end{equation}
where agent $i$ prefers choice $j_1$ to choice $j_2$ about $K$ times, and $\Delta$ is a constant that describes the degree of uncertainty in the accuracy of the parameter $K$.
This is especially useful when $\{q_{ij}\}_{j\in C}$ are known up to a multiplicative constant following estimation via questionnaires and surveys (these, in some sense, ``isolate'' agents from their network). Typically, this should provide $|C|(|C|-1)/2$ pairs of inequality constraints per agent, or $|C|-1$ equality constraints when $\Delta=0$.

	\ \\
	\noindent{\em Decisiveness.} To represent the relative reliance by an agent on his/her network:
\begin{equation}\label{estimation_constraints_additional_info_3}
	\sum_{k\in A}{p_{ik}} \ge K \sum_{j\in C}{q_{ij}}
	{\rm\ \ or\ \ }
	\sum_{k\in A}{p_{ik}} \le K \sum_{j\in C}{q_{ij}}
\end{equation}
for some $K$.
	
	\ \\
	\noindent{\em Presence/absence of reliance on other agents (sparsity).} Where it is clear or highly likely that, a given agent $i$ never directly adopts the choice of another agent $k$, one may represent this as
\begin{equation}\label{estimation_constraints_additional_info_4}
	p_{ik} = 0.
\end{equation}
This is particularly important because without constraints of this sort, the dimension of ``real-world problems'' would become too large to be computationally tractable in practice.

\subsection*{Adjusting the Polyhedron so That It Is Nonempty}
	As a preprocessing step, to ensure that the polyhedron to be searched for a point estimate of the model parameters is non-empty, one might solve a linear optimization problem comprising constraints (\ref{estimation_constraints_1})-(\ref{estimation_constraints_4}) as well as constraints of the form (\ref{estimation_constraints_additional_info_1})-(\ref{estimation_constraints_additional_info_4}).
The goal of doing this is to find the ``smallest'' slack variables necessary for feasibility.
To this end, multiple objectives are possible, for example, one might minimize the largest ratio of $\epsilon_{ij}^{+}+\epsilon_{ij}^{-}$ to $\pi_{ij}$. Once the foregoing optimization problem is solved, one obtains a polyhedron described by constraints (\ref{estimation_constraints_1})-(\ref{estimation_constraints_4}) (albeit with the $\epsilon_{ij}^{\cdot}$'s fixed), and the same constraints of the form (\ref{estimation_constraints_additional_info_1})-(\ref{estimation_constraints_additional_info_4}). Denote this polyhedron $S$.

\subsection*{Computing an Estimate for $\{p_{ik}\}_{i\in A, k\in A}$ and $\{q^{(j)}\}_{j\in C}$}
	
	At this point, one may finally solve an optimization problem to maximize, over $\{p_{ik}\}_{i\in A, k\in A}$ and $\{q^{(j)}\}_{j\in C}$ in $S$, the distance to the planes defined by the inequality constraints (\ref{estimation_constraints_additional_info_1})-(\ref{estimation_constraints_additional_info_3}). Should the optimal value of this problem be $0$, the constraints which are satisfied with equality for all points of $S$ should be replaced with equality constraints and the resulting problem solved. This should be repeated until either no inequality constraints are left, or a strictly positive optimal value is obtained.

	In either case, the final solution provides a point estimate. However, in the latter case, one has a point in the relative interior which may be used as a feasible starting point to solve for the analytic center (of the relative interior).


\section{More on Decisiveness}\label{appendix_ex_decisiveness}

	Consider an example with $C = \{-1, 1\}$ and $q_{i,j} = \eta/2$ for both $j\in C$. Suppose also that $P$ is a matrix with equal entries in all its off diagonal entries and zeros on its diagonal.
	In this case, it is clear that the choice shares are equal.
	Under the simulation process in \textsection\ref{ss_comparison_KKT}, each non-rejected realization yields a ``choice vector'' $X \in C^{|A|} = \{-1,1\}^{|A|}$ listing the respective choices of each agent.
	Obviously, $\Ex{X} = 0$ since the choice shares are equal. However, one may easily verify that $\lim\limits_{\eta\downarrow0}{\Ex{XX^T} = ee^{T}}$. This occurs because it becomes increasingly unlikely (with decreasing $\eta$) that more than one agent selects a choice without relying on others in the network.

	With the above example in mind, it is clear that in regimes of low decisiveness, as quantified by by $\{\sum_{j\in C}{q_{ij}}\}_{i\in A}$, herding emerges. This is explored briefly in a special case.

\subsection{Herding by Preferential Attachment}	

	Consider $P$ such that for each $i\in A$, $p_{i k_1} = p_{i k_2}$ for all $k_1,k_2 \in A$ ($k_1\not=k_2\not=i$). It may be deduced that conditional on the ``decisive'' agents (those that selected choices from $C$ in the first step of the simulation process of \textsection\ref{ss_comparison_KKT}), each other agent adopts the choice of some decisive agent (via a chain of adoptions) according to a proportional preferential attachment process. (This is also known as the Polya urn model or, more generally, a finite Polya process with ``unit feedback''. See, for instance \cite{Chung2003}.)

	Each decisive agent corresponds to a set of agents that ultimately adopt his/her choice via a chain of agents. Let each such set be known as a herd. Clearly, the number of herds is a random variable that depends on $\{\sum_{j\in C}{q_{ij}}\}_{i\in A}$, with the expected number of herds monotonically increasing in it.

	For this special case, Proposition \ref{prop_max_herd_size}
(stated without proof)
gives the expected fraction of agents in the largest herd conditional on the number of decisive agents. The result is asymptotic due to ``edge-effects'' associated with the discreteness of outcomes. However, simulation results suggest that the limit distribution is a good approximation even for populations of agents that one might not regard as large. Denote the maximum herd size when there are $d$ decisive agents among a total of $|A|$ agents as $H_d^{\max}(|A|)$. 



\begin{proposition}[Expected Fraction of Agents in the Largest Herd]\label{prop_max_herd_size}
	\begin{equation}\label{asymp_max_herd_size}
		\lim_{|A|\rightarrow\infty}\Ex{\frac{H_d^{\max}(|A|)}{|A|}} = \frac{1}{d}\sum_{k=1}^{d}{\frac{1}{k}} =: M^{(1)}_d.
	\end{equation}
\end{proposition}

	Illustrations of Proposition \ref{prop_max_herd_size} may be found in Figures \ref{fig_asymp_max_herd_size_100} thru \ref{fig_asymp_max_herd_size_20000} along with simulated average maximum herd sizes and quantiles obtained from simulation (for $|A|$ ranging from 100 to 20,000).
Note the similarity of the figures over the large range of $|A|$.


	\definecolor{orange}{rgb}{1,0.5,0}
	\definecolor{brown}{rgb}{0.75,0.25,0}

	\begin{figure}[H]\begin{center}
		\begin{tikzpicture}[y=4.05cm, x=.485cm]
			\draw (0,0) -- coordinate (x axis mid) (25,0);
		    	\draw (0,0) -- coordinate (y axis mid) (0,1);
		    	\foreach \x in {0,5,...,25} {
		     		\draw (\x,3pt) -- (\x,-3pt)
					node[anchor=north] {\x};
				}
		    	\foreach \y in {0,0.2,0.4,0.6,0.8,1} {
					\draw (3pt,\y) -- (-3pt,\y) 
		     			node[anchor=east] {\y}; 
				}
			\node[below=0.4cm] at (x axis mid) {$d$};
			\node[rotate=90, above=0.8cm] at (y axis mid) {$\displaystyle{H_d^{\max}(|A|)} / |A|$};
		
			\draw[blue, thick] plot file {herding_asymp_exact_25.data};
			\draw[red, thick] plot[mark=*, only marks] file {herding_sim_100.data};
			\draw[brown, dashed] plot file {herding_sim_100_q20.data};
			\draw[brown, dashed] plot file {herding_sim_100_q80.data};
			\draw[orange, dashdotted] plot file {herding_sim_100_q05.data};
			\draw[orange, dashdotted] plot file {herding_sim_100_q95.data};

			\begin{scope}[shift={(9,0.985)}]
				\draw[blue, thick, yshift=0\baselineskip]
					plot[blue] (0,0) -- (1.5,0) 
					node[right]{Expectation: Proposition \ref{prop_max_herd_size} ($|A|\rightarrow\infty$)};
				\draw[red, thick, yshift=-1\baselineskip]
					plot[mark=*, only marks] (0.75,0);
				\draw[red, thick, yshift=-1\baselineskip] (1.5,0) -- (1.5,0) 
					node[right]{Simulation Mean};
				\draw[brown, dashed, yshift=-2\baselineskip] (0,0) -- 
					plot[brown] (0,0) -- (1.5,0) 
					node[right]{Simulation Quantiles: 0.2 and 0.8};
				\draw[orange, dashdotted, yshift=-3\baselineskip] (0,0) -- 
					plot[orange] (0,0) -- (1.5,0) 
					node[right]{Simulation Quantiles: 0.05 and 0.95};
			\end{scope}
		\end{tikzpicture}
		\caption{Expected Maximum Herd Size: The Asymptotic Limit and Simulated Data ($|A|={\rm 100}$; 10,000 samples)}
		\label{fig_asymp_max_herd_size_100}
	\end{center}\end{figure}

	\begin{figure}[H]\begin{center}
		\begin{tikzpicture}[y=4.05cm, x=.485cm]
			\draw (0,0) -- coordinate (x axis mid) (25,0);
		    	\draw (0,0) -- coordinate (y axis mid) (0,1);
		    	\foreach \x in {0,5,...,25} {
		     		\draw (\x,3pt) -- (\x,-3pt)
					node[anchor=north] {\x};
				}
		    	\foreach \y in {0,0.2,0.4,0.6,0.8,1} {
					\draw (3pt,\y) -- (-3pt,\y) 
		     			node[anchor=east] {\y}; 
				}
			\node[below=0.4cm] at (x axis mid) {$d$};
			\node[rotate=90, above=0.8cm] at (y axis mid) {$\displaystyle{H_d^{\max}(|A|)} / |A|$};
		
			\draw[blue, thick] plot file {herding_asymp_exact_25.data};
			\draw[red, thick] plot[mark=*, only marks] file {herding_sim_250.data};
			\draw[brown, dashed] plot file {herding_sim_250_q20.data};
			\draw[brown, dashed] plot file {herding_sim_250_q80.data};
			\draw[orange, dashdotted] plot file {herding_sim_250_q05.data};
			\draw[orange, dashdotted] plot file {herding_sim_250_q95.data};

			\begin{scope}[shift={(9,0.985)}]
				\draw[blue, thick, yshift=0\baselineskip]
					plot[blue] (0,0) -- (1.5,0) 
					node[right]{Expectation: Proposition \ref{prop_max_herd_size} ($|A|\rightarrow\infty$)};
				\draw[red, thick, yshift=-1\baselineskip]
					plot[mark=*, only marks] (0.75,0);
				\draw[red, thick, yshift=-1\baselineskip] (1.5,0) -- (1.5,0) 
					node[right]{Simulation Mean};
				\draw[brown, dashed, yshift=-2\baselineskip] (0,0) -- 
					plot[brown] (0,0) -- (1.5,0) 
					node[right]{Simulation Quantiles: 0.2 and 0.8};
				\draw[orange, dashdotted, yshift=-3\baselineskip] (0,0) -- 
					plot[orange] (0,0) -- (1.5,0) 
					node[right]{Simulation Quantiles: 0.05 and 0.95};
			\end{scope}
		\end{tikzpicture}
		\caption{Expected Maximum Herd Size: The Asymptotic Limit and Simulated Data ($|A|={\rm 250}$; 10,000 samples)}
		\label{fig_asymp_max_herd_size_250}
	\end{center}\end{figure}

	\begin{figure}[H]\begin{center}
		\begin{tikzpicture}[y=4.05cm, x=.485cm]
			\draw (0,0) -- coordinate (x axis mid) (25,0);
		    	\draw (0,0) -- coordinate (y axis mid) (0,1);
		    	\foreach \x in {0,5,...,25} {
		     		\draw (\x,3pt) -- (\x,-3pt)
					node[anchor=north] {\x};
				}
		    	\foreach \y in {0,0.2,0.4,0.6,0.8,1} {
					\draw (3pt,\y) -- (-3pt,\y) 
		     			node[anchor=east] {\y}; 
				}
			\node[below=0.4cm] at (x axis mid) {$d$};
			\node[rotate=90, above=0.8cm] at (y axis mid) {$\displaystyle{H_d^{\max}(|A|)} / |A|$};
		
			\draw[blue, thick] plot file {herding_asymp_exact_25.data};
			\draw[red, thick] plot[mark=*, only marks] file {herding_sim_500.data};
			\draw[brown, dashed] plot file {herding_sim_500_q20.data};
			\draw[brown, dashed] plot file {herding_sim_500_q80.data};
			\draw[orange, dashdotted] plot file {herding_sim_500_q05.data};
			\draw[orange, dashdotted] plot file {herding_sim_500_q95.data};

			\begin{scope}[shift={(9,0.985)}]
				\draw[blue, thick, yshift=0\baselineskip]
					plot[blue] (0,0) -- (1.5,0) 
					node[right]{Expectation: Proposition \ref{prop_max_herd_size} ($|A|\rightarrow\infty$)};
				\draw[red, thick, yshift=-1\baselineskip]
					plot[mark=*, only marks] (0.75,0);
				\draw[red, thick, yshift=-1\baselineskip] (1.5,0) -- (1.5,0) 
					node[right]{Simulation Mean};
				\draw[brown, dashed, yshift=-2\baselineskip] (0,0) -- 
					plot[brown] (0,0) -- (1.5,0) 
					node[right]{Simulation Quantiles: 0.2 and 0.8};
				\draw[orange, dashdotted, yshift=-3\baselineskip] (0,0) -- 
					plot[orange] (0,0) -- (1.5,0) 
					node[right]{Simulation Quantiles: 0.05 and 0.95};
			\end{scope}
		\end{tikzpicture}
		\caption{Expected Maximum Herd Size: The Asymptotic Limit and Simulated Data ($|A|={\rm 500}$; 10,000 samples)}
		\label{fig_asymp_max_herd_size_500}
	\end{center}\end{figure}

	\begin{figure}[H]\begin{center}
		\begin{tikzpicture}[y=4.05cm, x=.485cm]
			\draw (0,0) -- coordinate (x axis mid) (25,0);
		    	\draw (0,0) -- coordinate (y axis mid) (0,1);
		    	\foreach \x in {0,5,...,25} {
		     		\draw (\x,3pt) -- (\x,-3pt)
					node[anchor=north] {\x};
				}
		    	\foreach \y in {0,0.2,0.4,0.6,0.8,1} {
					\draw (3pt,\y) -- (-3pt,\y) 
		     			node[anchor=east] {\y}; 
				}
			\node[below=0.4cm] at (x axis mid) {$d$};
			\node[rotate=90, above=0.8cm] at (y axis mid) {$\displaystyle{H_d^{\max}(|A|)} / |A|$};
		
			\draw[blue, thick] plot file {herding_asymp_exact_25.data};
			\draw[red, thick] plot[mark=*, only marks] file {herding_sim_1000.data};
			\draw[brown, dashed] plot file {herding_sim_1000_q20.data};
			\draw[brown, dashed] plot file {herding_sim_1000_q80.data};
			\draw[orange, dashdotted] plot file {herding_sim_1000_q05.data};
			\draw[orange, dashdotted] plot file {herding_sim_1000_q95.data};

			\begin{scope}[shift={(9,0.985)}]
				\draw[blue, thick, yshift=0\baselineskip]
					plot[blue] (0,0) -- (1.5,0) 
					node[right]{Expectation: Proposition \ref{prop_max_herd_size} ($|A|\rightarrow\infty$)};
				\draw[red, thick, yshift=-1\baselineskip]
					plot[mark=*, only marks] (0.75,0);
				\draw[red, thick, yshift=-1\baselineskip] (1.5,0) -- (1.5,0) 
					node[right]{Simulation Mean};
				\draw[brown, dashed, yshift=-2\baselineskip] (0,0) -- 
					plot[brown] (0,0) -- (1.5,0) 
					node[right]{Simulation Quantiles: 0.2 and 0.8};
				\draw[orange, dashdotted, yshift=-3\baselineskip] (0,0) -- 
					plot[orange] (0,0) -- (1.5,0) 
					node[right]{Simulation Quantiles: 0.05 and 0.95};
			\end{scope}
		\end{tikzpicture}
		\caption{Expected Maximum Herd Size: The Asymptotic Limit and Simulated Data ($|A|={\rm 1,000}$; 10,000 samples)}
		\label{fig_asymp_max_herd_size_1000}
	\end{center}\end{figure}

	\begin{figure}[H]\begin{center}
		\begin{tikzpicture}[y=4.05cm, x=.485cm]
			\draw (0,0) -- coordinate (x axis mid) (25,0);
		    	\draw (0,0) -- coordinate (y axis mid) (0,1);
		    	\foreach \x in {0,5,...,25} {
		     		\draw (\x,3pt) -- (\x,-3pt)
					node[anchor=north] {\x};
				}
		    	\foreach \y in {0,0.2,0.4,0.6,0.8,1} {
					\draw (3pt,\y) -- (-3pt,\y) 
		     			node[anchor=east] {\y}; 
				}
			\node[below=0.4cm] at (x axis mid) {$d$};
			\node[rotate=90, above=0.8cm] at (y axis mid) {$\displaystyle{H_d^{\max}(|A|)} / |A|$};
		
			\draw[blue, thick] plot file {herding_asymp_exact_25.data};
			\draw[red, thick] plot[mark=*, only marks] file {herding_sim_2000.data};
			\draw[brown, dashed] plot file {herding_sim_2000_q20.data};
			\draw[brown, dashed] plot file {herding_sim_2000_q80.data};
			\draw[orange, dashdotted] plot file {herding_sim_2000_q05.data};
			\draw[orange, dashdotted] plot file {herding_sim_2000_q95.data};

			\begin{scope}[shift={(9,0.985)}]
				\draw[blue, thick, yshift=0\baselineskip]
					plot[blue] (0,0) -- (1.5,0) 
					node[right]{Expectation: Proposition \ref{prop_max_herd_size} ($|A|\rightarrow\infty$)};
				\draw[red, thick, yshift=-1\baselineskip]
					plot[mark=*, only marks] (0.75,0);
				\draw[red, thick, yshift=-1\baselineskip] (1.5,0) -- (1.5,0) 
					node[right]{Simulation Mean};
				\draw[brown, dashed, yshift=-2\baselineskip] (0,0) -- 
					plot[brown] (0,0) -- (1.5,0) 
					node[right]{Simulation Quantiles: 0.2 and 0.8};
				\draw[orange, dashdotted, yshift=-3\baselineskip] (0,0) -- 
					plot[orange] (0,0) -- (1.5,0) 
					node[right]{Simulation Quantiles: 0.05 and 0.95};
			\end{scope}
		\end{tikzpicture}
		\caption{Expected Maximum Herd Size: The Asymptotic Limit and Simulated Data ($|A|={\rm 2,000}$; 10,000 samples)}
		\label{fig_asymp_max_herd_size_2000}
	\end{center}\end{figure}

	\begin{figure}[H]\begin{center}
		\begin{tikzpicture}[y=4.05cm, x=.485cm]
			\draw (0,0) -- coordinate (x axis mid) (25,0);
		    	\draw (0,0) -- coordinate (y axis mid) (0,1);
		    	\foreach \x in {0,5,...,25} {
		     		\draw (\x,3pt) -- (\x,-3pt)
					node[anchor=north] {\x};
				}
		    	\foreach \y in {0,0.2,0.4,0.6,0.8,1} {
					\draw (3pt,\y) -- (-3pt,\y) 
		     			node[anchor=east] {\y}; 
				}
			\node[below=0.4cm] at (x axis mid) {$d$};
			\node[rotate=90, above=0.8cm] at (y axis mid) {$\displaystyle{H_d^{\max}(|A|)} / |A|$};
		
			\draw[blue, thick] plot file {herding_asymp_exact_25.data};
			\draw[red, thick] plot[mark=*, only marks] file {herding_sim_5000.data};
			\draw[brown, dashed] plot file {herding_sim_5000_q20.data};
			\draw[brown, dashed] plot file {herding_sim_5000_q80.data};
			\draw[orange, dashdotted] plot file {herding_sim_5000_q05.data};
			\draw[orange, dashdotted] plot file {herding_sim_5000_q95.data};

			\begin{scope}[shift={(9,0.985)}]
				\draw[blue, thick, yshift=0\baselineskip]
					plot[blue] (0,0) -- (1.5,0) 
					node[right]{Expectation: Proposition \ref{prop_max_herd_size} ($|A|\rightarrow\infty$)};
				\draw[red, thick, yshift=-1\baselineskip]
					plot[mark=*, only marks] (0.75,0);
				\draw[red, thick, yshift=-1\baselineskip] (1.5,0) -- (1.5,0) 
					node[right]{Simulation Mean};
				\draw[brown, dashed, yshift=-2\baselineskip] (0,0) -- 
					plot[brown] (0,0) -- (1.5,0) 
					node[right]{Simulation Quantiles: 0.2 and 0.8};
				\draw[orange, dashdotted, yshift=-3\baselineskip] (0,0) -- 
					plot[orange] (0,0) -- (1.5,0) 
					node[right]{Simulation Quantiles: 0.05 and 0.95};
			\end{scope}
		\end{tikzpicture}
		\caption{Expected Maximum Herd Size: The Asymptotic Limit and Simulated Data ($|A|={\rm 5,000}$; 10,000 samples)}
		\label{fig_asymp_max_herd_size_2000}
	\end{center}\end{figure}

	\begin{figure}[H]\begin{center}
		\begin{tikzpicture}[y=4.05cm, x=.485cm]
			\draw (0,0) -- coordinate (x axis mid) (25,0);
		    	\draw (0,0) -- coordinate (y axis mid) (0,1);
		    	\foreach \x in {0,5,...,25} {
		     		\draw (\x,3pt) -- (\x,-3pt)
					node[anchor=north] {\x};
				}
		    	\foreach \y in {0,0.2,0.4,0.6,0.8,1} {
					\draw (3pt,\y) -- (-3pt,\y) 
		     			node[anchor=east] {\y}; 
				}
			\node[below=0.4cm] at (x axis mid) {$d$};
			\node[rotate=90, above=0.8cm] at (y axis mid) {$\displaystyle{H_d^{\max}(|A|)} / |A|$};
		
			\draw[blue, thick] plot file {herding_asymp_exact_25.data};
			\draw[red, thick] plot[mark=*, only marks] file {herding_sim_10000.data};
			\draw[brown, dashed] plot file {herding_sim_10000_q20.data};
			\draw[brown, dashed] plot file {herding_sim_10000_q80.data};
			\draw[orange, dashdotted] plot file {herding_sim_10000_q05.data};
			\draw[orange, dashdotted] plot file {herding_sim_10000_q95.data};

			\begin{scope}[shift={(9,0.985)}]
				\draw[blue, thick, yshift=0\baselineskip]
					plot[blue] (0,0) -- (1.5,0) 
					node[right]{Expectation: Proposition \ref{prop_max_herd_size} ($|A|\rightarrow\infty$)};
				\draw[red, thick, yshift=-1\baselineskip]
					plot[mark=*, only marks] (0.75,0);
				\draw[red, thick, yshift=-1\baselineskip] (1.5,0) -- (1.5,0) 
					node[right]{Simulation Mean};
				\draw[brown, dashed, yshift=-2\baselineskip] (0,0) -- 
					plot[brown] (0,0) -- (1.5,0) 
					node[right]{Simulation Quantiles: 0.2 and 0.8};
				\draw[orange, dashdotted, yshift=-3\baselineskip] (0,0) -- 
					plot[orange] (0,0) -- (1.5,0) 
					node[right]{Simulation Quantiles: 0.05 and 0.95};
			\end{scope}
		\end{tikzpicture}
		\caption{Expected Maximum Herd Size: The Asymptotic Limit and Simulated Data ($|A|={\rm 10,000}$; 10,000 samples)}
		\label{fig_asymp_max_herd_size_10000}
	\end{center}\end{figure}

	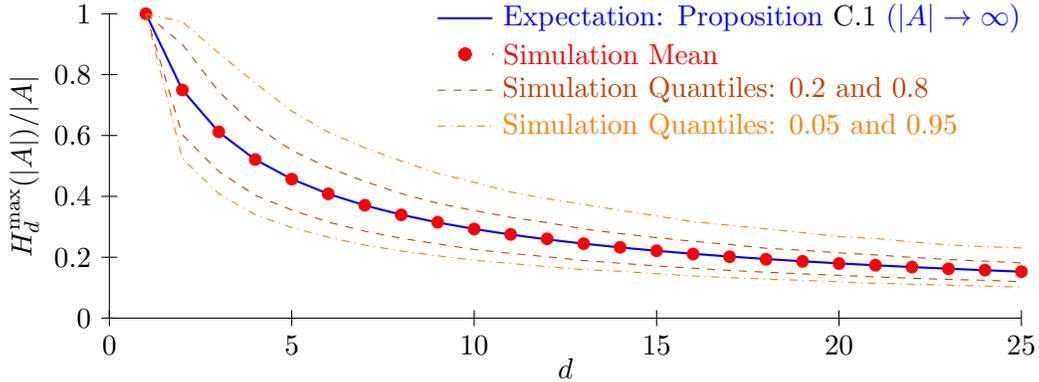
\begin{figure}[H]\begin{center}
		\begin{tikzpicture}[y=4.05cm, x=.485cm]
			\draw (0,0) -- coordinate (x axis mid) (25,0);
		    	\draw (0,0) -- coordinate (y axis mid) (0,1);
		    	\foreach \x in {0,5,...,25} {
		     		\draw (\x,3pt) -- (\x,-3pt)
					node[anchor=north] {\x};
				}
		    	\foreach \y in {0,0.2,0.4,0.6,0.8,1} {
					\draw (3pt,\y) -- (-3pt,\y) 
		     			node[anchor=east] {\y}; 
				}
			\node[below=0.4cm] at (x axis mid) {$d$};
			\node[rotate=90, above=0.8cm] at (y axis mid) {$\displaystyle{H_d^{\max}(|A|)} / |A|$};
		
			\draw[blue, thick] plot file {herding_asymp_exact_25.data};
			\draw[red, thick] plot[mark=*, only marks] file {herding_sim_20000.data};
			\draw[brown, dashed] plot file {herding_sim_20000_q20.data};
			\draw[brown, dashed] plot file {herding_sim_20000_q80.data};
			\draw[orange, dashdotted] plot file {herding_sim_20000_q05.data};
			\draw[orange, dashdotted] plot file {herding_sim_20000_q95.data};

			\begin{scope}[shift={(9,0.985)}]
				\draw[blue, thick, yshift=0\baselineskip]
					plot[blue] (0,0) -- (1.5,0) 
					node[right]{Expectation: Proposition \ref{prop_max_herd_size} ($|A|\rightarrow\infty$)};
				\draw[red, thick, yshift=-1\baselineskip]
					plot[mark=*, only marks] (0.75,0);
				\draw[red, thick, yshift=-1\baselineskip] (1.5,0) -- (1.5,0) 
					node[right]{Simulation Mean};
				\draw[brown, dashed, yshift=-2\baselineskip] (0,0) -- 
					plot[brown] (0,0) -- (1.5,0) 
					node[right]{Simulation Quantiles: 0.2 and 0.8};
				\draw[orange, dashdotted, yshift=-3\baselineskip] (0,0) -- 
					plot[orange] (0,0) -- (1.5,0) 
					node[right]{Simulation Quantiles: 0.05 and 0.95};
			\end{scope}
		\end{tikzpicture}
		\caption{Expected Maximum Herd Size: The Asymptotic Limit and Simulated Data ($|A|={\rm 20,000}$; 10,000 samples)}
		\label{fig_asymp_max_herd_size_20000}
	\end{center}\end{figure}

	To frame this result in a more classical, combinatorial context, consider finite Polya processes with $d$ ``bins'' in which the probability that the next ``ball'' is placed in a ``bin'' is proportional to $m^\gamma$ where $m$ is the number of ``balls'' currently in that ``bin''. It is well known that when $\gamma < 1$, the limiting fraction of ``balls'' in the ``largest'' ``bin'' is $1/d$. When $\gamma > 1$, the limiting fraction is $1$. Proposition \ref{prop_max_herd_size} (and its generalization, Proposition \ref{prop_max_herd_size_moments}) fills in a small part of the gap for the case where $\gamma = 1$. (Note: The paper of \cite{Chung2003} includes a section containing a short survey of known results for finite Polya processes.)

	Under this stylized model, Proposition \ref{prop_max_herd_size} suggests that herding behavior is of concern mainly in environments with few ``decisive'' agents where the largest herd can correspond to a relatively large fraction of agents. While it would be inappropriate to draw too many conclusions from this stylized model, however it is somewhat telling that \cite{Barabasi2003} found that
``{\it measurements on four networks, namely the science citation network, Internet, actor collaboration and science coauthorship network indicate that the rate at which nodes acquire links depends on the node's degree, offering direct quantitative support for the presence of preferential attachment}''. (Admittedly, citations and Internet links are compatible with ``choice'', while the others reflect ``group formation''.)

	As a point of note, the proof technique of Proposition \ref{prop_max_herd_size} may also be used to obtain all the higher moments for the fraction of the total population in the largest herd (result stated without proof):
\begin{proposition}[Moments of the Fraction of Agents in the Largest Herd]\label{prop_max_herd_size_moments}
	Let
	$$
		M^{(m)}_d := \lim_{|A|\rightarrow\infty}\Ex{\left(\frac{H_d^{\max}(|A|)}{|A|}\right)^m}.
	$$
	for $d\ge 2$ and $m\ge 1$.
	The moments of the expected fraction of agents in the largest herd satisfy
	\begin{equation}\label{asymp_frac_moments_1}
		M^{(m)}_d
		= \sum_{k=0}^{m}{
			\frac{d-1}{d^k} 
			\frac{m!}{(m-k)!}
			\frac{(m+d-k-2)!}{(m+d-1)!}
			M^{(m-k)}_{d-1}
		},
	\end{equation}
	\begin{equation}\label{asymp_frac_moments_2}
		M^{(m)}_d
		= \frac{d-1}{m+d-1} M^{(m)}_{d-1} + \frac{m}{d(m+d-1)} M^{(m-1)}_d, {\rm\ and}
	\end{equation}
	\begin{equation}\label{asymp_frac_moments_3}
		M^{(m)}_d
		= \frac{m}{m+d-1} \sum_{j=1}^{d}{ \frac{(d-1)!}{j!} \frac{(m+j-2)!}{(m+d-2)!} M^{(m-1)}_j}.
	\end{equation}
	where $M^{(0)}_0 := 1$. In particular, $M^{(m)}_1 := 1$ and $M^{(0)}_d := 1$.
\end{proposition}

Simulations show that those limiting values are good approximations even for systems with just thousands of agents. As such, if we further restrict ourselves to the case what each agent is decisive with the same probability ($P = \frac{1-\gamma}{|A|}ee^{T})$ for $\gamma\in(0,1)$), one may efficiently compute approximate expectations of (smooth) functions of the fraction of the total population in the largest herd. To do this, one uses Proposition \ref{prop_max_herd_size_moments} to obtain a truncated series of moments given that number of decisive agents and uses an appropriate normal distribution to approximate binomial distribution with $|A|$ trials and parameter $\gamma$. The aforementioned normal approximation is used to sample the number of decisive agents and the moments are used to approximate the aforementioned functions.


\section{More on the Brand Ambassador Problem}\label{appendix_more_BA}

	The brand ambassador selection problem may be easily extended to consider situations where brand ambassadors never exclusively recommend a choice, such as in social media settings. This is achieved by ``splitting'' agents into multiple identical ``sub-agents'' who divide the original agents' endowment and may be ``individually'' added to the set of brand ambassadors. To achieve non-exclusive recommendation, one may disallow certain ``sub-agents'' from being added to the set of brand ambassadors. Naturally, this variant admits the same approximability result.

	In some problem settings, it may turn out that the ``network'' is revealed sequentially as a (random) function of the ``next'' agent added to the set of brand ambassadors. In this case, one may obtain the same approximation guarantee by recourse to the notions of adaptive monotonicity and adaptive submodularity, introduced by \cite{Golovin2011}, which is a generalization of submodularity.

	In settings with strategic agents who announce a cost of turning them into brand ambassadors, \cite{Singer2012} showed that for ``monotone submodular influence models'' there exists randomized selection rules (mechanisms) that would also ensure truthfulness in agents' reports of their costs (truthfulness), that selections respect budget constraints, and that outcomes are within a constant factor of optimality in expectation. In particular, he describes an incentive compatible mechanism for the ``coverage model'' wherein agents are greedily selected according to their marginal contribution per unit cost.
However, in contrast with the work of \cite{Sviridenko2004}, who described a variant of the greedy selection strategy for monotone submodular maximization that produces a $\left(1 - \frac{1}{e}\right)$ approximation of the optimal solution given knapsack/budget constraints, the strength of the approximation guarantees of \cite{Singer2012} are sacrificed in favor of computational tractability and ease of describing the mechanism.



\bibliographystyle{ormsv080} 
\bibliography{bib}		
